\documentclass{WileyMSP-template}
\usepackage[utf8]{inputenc}

\usepackage{bm} 
\usepackage{amsmath}
\usepackage{amssymb}
\usepackage{textcomp, gensymb}
\usepackage{siunitx}
\usepackage[version=4]{mhchem}
\usepackage{pdflscape}
\usepackage{cite}

\usepackage{threeparttable}

\newcommand{\tens}[1]{\boldsymbol{#1}}
\usepackage[capitalise,noabbrev]{cleveref}

\begin{document}

\pagestyle{fancy}
\rhead{\includegraphics[width=2.5cm]{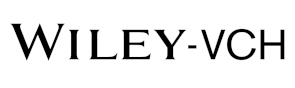}}

\title{Electric-Field-Induced Second-Harmonic Generation}

\maketitle


\author{Hangkai Fan,}
\author{Alexey Proskurin,}
\author{Mingzhao Song*,}
\author{Yuri Kivshar*,}
\author{Andrey Bogdanov*}



\begin{affiliations}
Dr. Hangkai Fan\\
Qingdao Innovation and Development Center, Harbin Engineering University, Qingdao 266000, Shandong, China.\\
Key Laboratory of Photonic Materials and Device Physics for Oceanic Applications, Ministry of Industry and Information Technology of China\\
College of information and communication engineering, Harbin Engineering University\\
School of Physics and Engineering, ITMO University, St. Petersburg 191002, Russia.\\

Dr. Alexey Proskurin\\
School of Physics and Engineering, ITMO University, St. Petersburg 191002, Russia.\\

Prof. Mingzhao Song\\
Qingdao Innovation and Development Center, Harbin Engineering University, Qingdao 266000, Shandong, China.\\
Key Laboratory of Photonic Materials and Device Physics for Oceanic Applications, Ministry of Industry and Information Technology of China\\
College of Physics and Optoelectronic Engineering, Harbin Engineering University\\
Email Address: kevinsmz@foxmail.com\\

Prof. Yuri Kivshar\\
Nonlinear Physics Center, Research School of Physics, Australian National University, Canberra ACT 2601, Australia.\\
Email Address: yuri.kivshar@anu.edu.au\\

Prof. Andrey Bogdanov\\
Qingdao Innovation and Development Center, Harbin Engineering University, Qingdao 266000, Shandong, China; 
School of Physics and Engineering, ITMO University, St. Petersburg 191002, Russia.\\
Email Address: bogdan.taurus@gmail.com

\end{affiliations}



\begin{abstract}
Second-harmonic generation (SHG) is a fundamental nonlinear optical process widely used in photonics; however, it is strictly forbidden in the bulk of centrosymmetric materials due to their inversion symmetry. Nevertheless, applying an external electric field breaks this inversion symmetry. It induces an effective second-order nonlinear response known as the electric-field-induced second-harmonic generation (EFISH) effect. This mechanism enables SHG even in centrosymmetric media and provides a powerful tool for dynamic and electrically tunable nonlinear nanophotonics. This review presents a comprehensive overview of the EFISH effect, covering the fundamentals, various material platforms (including bulk semiconductor crystals, ferroelectrics, van der Waals materials, and polymers), as well as diverse strategies for electric field engineering. It further distinguishs EFISH from related effects such as current-induced SHG and the quantum-confined Stark effect, and highlight emerging applications of EFISH in tunable photonic devices, carrier dynamics probing, and nonlinear optical modulation across optical, electronic, and THz regimes. Finally, key challenges and perspectives for the future development of electrically controlled nonlinear optical systems are outlined.

\end{abstract}


\section{Introduction}

Second harmonic generation (SHG) is a fundamental second-order nonlinear optical process that coherently doubles the frequency of incident light~\cite{boydNonlinearOptics2019}. The pioneering experimental demonstration by Franken et~al. in 1961~\cite{1961Phys.Rev.Lett.FrankenGenerationOpticalHarmonics} and the subsequent theoretical framework developed by Bloembergen et~al.~\cite{1962Phys.Rev.ArmstrongInteractionsLightWaves,1962Phys.Rev.BloembergenLightWavesBoundary} have established SHG as one of the cornerstones of modern optical technologies, including imaging, sensing, and quantum communication~\cite{krasnokNonlinearMetasurfacesParadigm2018,zhaoNonlinearMetaopticsApplications2021a,2024Nat.Photon.LuEmergingIntegratedLaser}. The mechanism behind SHG is the existence of second order nonlinear polarization, expressed as $\bm{P}_{2\omega} = \tens{\chi}^{(2)}:\bm{E}_{\omega}\otimes\bm{E}_{\omega}$ in frequency domain, where $\tens{\chi}^{(2)}$ is a third-rank tensor intrinsically linked to the symmetry properties of the medium~\cite{boydNonlinearOptics2019}, and $\bm{E}_{\omega}$ is the total local electric field at the fundamental frequency $\omega$. Resonance effects in structured media can greatly enhance the magnitude of $\bm{E}_{\omega}$.

However, most reported SHG are restricted to a limited set of non-centrosymmetric materials, such as lithium niobate (\ce{LiNbO3})~\cite{doi:10.1126/science.abj4396} and gallium phosphide arsenide (\ce{GaP_xAs_{1-x}})~\cite{2024NanophotonicsWangReviewGalliumPhosphidea}, which naturally possess a large bulk $\tens{\chi}^{(2)}\sim 30-70~\text{pm V}^{-1}$~\cite{choy1976accurate,schneider2018gallium}. In contrast, common materials such as silicon, which are compatible with CMOS fabrication technology, are centrosymmetric and, thus, inherently forbid the existence of $\tens{\chi}^{(2)}$ in bulk, suppressing SHG. 

Several strategies have been developed to enable SHG in centrosymmetric materials. One approach exploits surface contributions~\cite{1986Phys.Rev.BGuyot-SionnestGeneralConsiderationsOptical}, where the naturally broken inversion symmetry at material interfaces permits SHG. However, this effect is confined to only a few atomic layers, resulting in limited efficiency. Alternatively, bulk inversion symmetry can be broken by applying external inhomogeneous stress to induce lattice deformation, thereby generating a strain-induced second-order susceptibility~\cite{2006NatureJacobsenStrainedSiliconNew,2012NatureMaterCazzanelliSecondharmonicGenerationSilicon,2015AdvancedOpticalMaterialsSchrieverSecondOrderOpticalNonlinearity,2022AdvancedPhotonicsResearchZhaoSecondHarmonicGenerationStrained}.

Another efficient method involves applying a static electric field, which perturbs the electronic potential without altering the crystallographic lattice structure, thereby inducing an effective second-order nonlinearity given by $\tens{\chi}^{(2)}_{\text{eff}} = 3\tens{\chi}^{(3)}\colon\bm{E}_{\mathrm{DC}}$~\cite{1962Phys.Rev.BloembergenLightWavesBoundary,1962Phys.Rev.Lett.TerhuneOpticalHarmonicGeneration}. This EFISH effect not only provides a method to induced SHG in centrosymmetry, but also provides a promising pathway for dynamic and reversible control of SHG through electrical modulation. Similarly to the linear Kerr effect, where the effective linear susceptibility is modulated as $\delta\tens{\chi}^{(1)}_{\text{eff}} \propto \tens{\chi}^{(3)}\colon(\bm{E}_{\mathrm{DC}}\otimes\bm{E}_\text{DC})$~\cite{soref1987electrooptical,2023AdvancedOpticalMaterialsKangNonlinearChiralMetasurfaces}, the EFISH mechanism enables modulation of the effective second-order susceptibility $\tens{\chi}^{(2)}_{\text{eff}}\propto\tens{\chi}^{(3)}\colon\bm{E}_{\mathrm{DC}}$~\cite{2019LightSciApplChenGiganticElectricfieldinducedSecond}, providing a promising approach for the development of electrically tunable nonlinear optical devices.

The application of a static electric field breaks the inversion symmetry of a centrosymmetric material, thereby enabling second-harmonic generation (SHG). However, the internal physical mechanisms responsible for symmetry breaking can vary significantly depending on the system. Therefore, it is important to distinguish conventional electric-field-induced second-harmonic generation (EFISH) from other related effects, such as current-induced second harmonic (CISH)~\cite{2010Phys.Rev.Lett.WangSecondOrderNonlinearOptical} and the quantum-confined Stark effect (QCSE)~\cite{1994IEEEJ.QuantumElectron.CapassoCoupledQuantumWell}. Although all of these effects allow electrical tuning of the effective second-order nonlinear susceptibility, their underlying mechanisms differ significantly. CISH occurs exclusively in semiconductors with high carrier densities, where substantial carrier migration, rather than the static electric field in the space-charge region, breaks the symmetry of electron distribution in momentum space, leading to SHG. QCSE is typically observed in low-dimensional structures such as van der Waals~(vdW) materials~\cite{2020ACSNanoDaiElectricalControlInterband,2016NanoLett.KleinStarkEffectSpectroscopy}, or multiple quantum wells~(MQW) ~\cite{1994IEEEJ.QuantumElectron.CapassoCoupledQuantumWell,2023AdvancedScienceChungElectricalPhaseModulation} due to their quantum confinement effect. It is the resonance behavior of $\tens{\chi}^{(2)}_\text{eff}$, originating from the strong coupling between the driving field and the electron wave equation (including both exciton formation and intersubband transitions~(IST) in MQWs), that underpins the giant nonlinear response. This resonance can be electrically tuned via the QCSE, which shifts the subband-exciton transition energies~\cite{2022Nat.Photon.YuElectricallyTunableNonlinear}, 
thereby enabling electrical control of resonance $\bm\chi^{(2)}_{\text{eff}}$.

\begin{figure}
  \includegraphics[width=\linewidth]{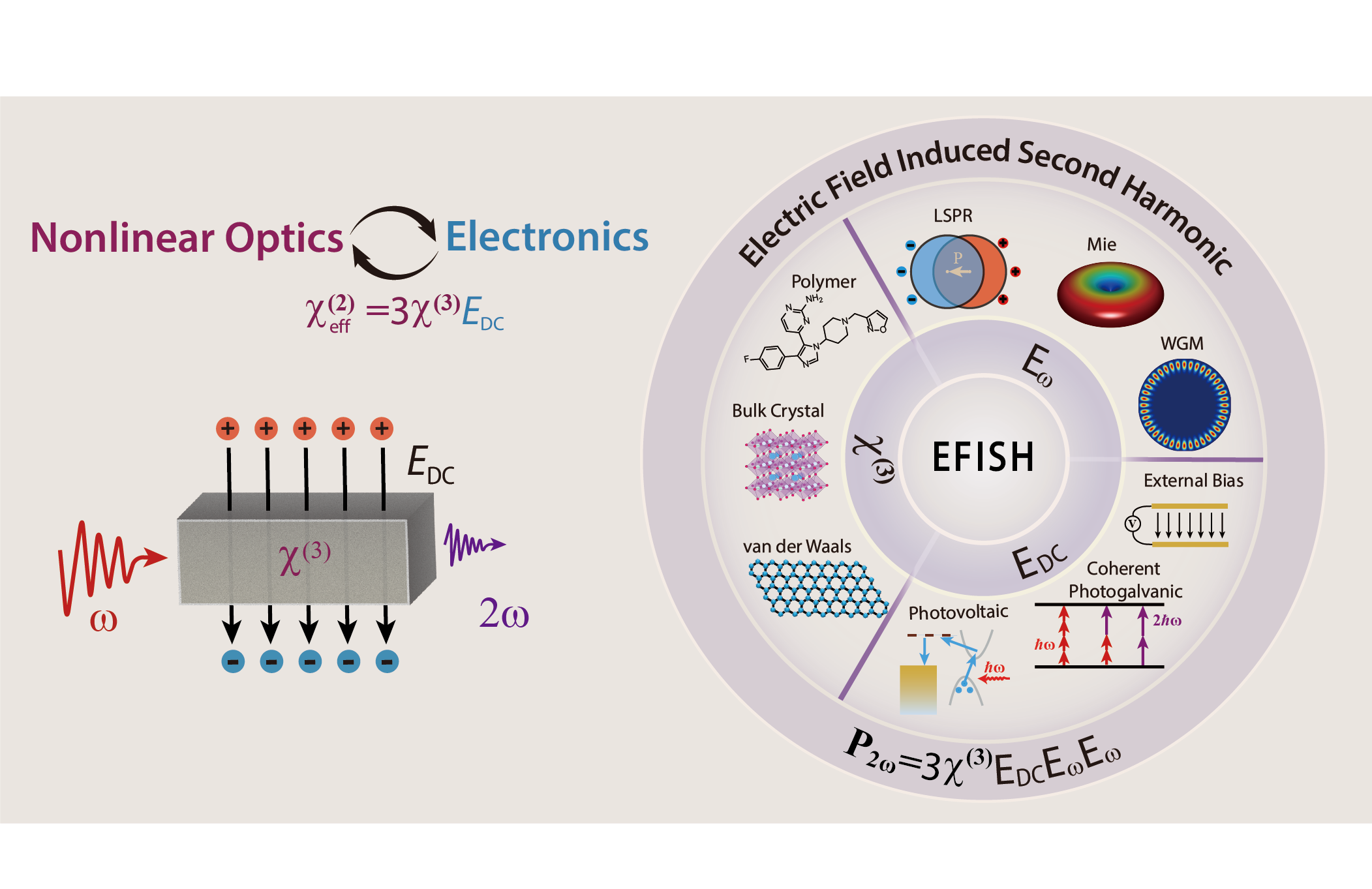}
  \caption{ \textbf{Schematic overview of the key topics covered in this review.} The EFISH effect emerges from the intimate coupling between electronic dynamics and nonlinear photonic. Depending on the EFISH functionality, the research can be categorized in to three topics:  (i)~nonlinear material engineering; (ii)~electrostatic field engineering; (iii)~fundamental electric field engineering.}
  \label{fig:boat1}
\end{figure}

In this work, we provide a comprehensive overview of recent advances in EFISH. \textbf{\cref{sec:fundamentals}} introduces the theoretical foundations of EFISH, laying the foundation for subsequent discussions. Recent progress in EFISH is reviewed in three key areas: (1)~nonlinear material selection, including bulk centrosymmetric crystals, ferroelectric materials, vdW materials, and organic polymers; (2)~engineering of static electric fields, covering both externally applied and optically generated fields; and (3)~photonics mode design, exploring EFISH enhancement through resonance effects in metallic and dielectric metasurfaces. \textbf{\cref{sec:CISH_QCSE}} provides a brief review of CISH and QCSE to highlight their distinct physical mechanisms and performance differences. \textbf{\cref{sec:applications}} addresses practical applications of EFISH, particularly in optical manipulation and carrier dynamics probing. Finally, we provide a comprehensive discussion on current challenges and promising future research directions. The key topics reviewed in this article are summarized schematically in \textbf{\cref{fig:boat1}}. 

\section{Fundamentals of EFISH}\label{sec:fundamentals}

When a photon impinges upon an electron in a material, the electron experiences a displacement disrupting electrical equilibrium, effectively forming an electric dipole. The resulting dipole oscillates generates a scattered electromagnetic field, which in turn modifies the total local field and gives rise to various optical phenomena such as reflection, transmission, and harmonic generation. Collectively, these induced dipoles constitute the polarization $\bm{P}$. Under dipole approximation, the optical polarization can be expanded as:

\begin{equation}
\bm{P}\;=\;
\tens{\chi}^{(1)} \colon \bm{E}\;+\;
\tens{\chi}^{(2)} \colon \bigl(\bm{E}\otimes\bm{E}\bigr)\;+\;
\tens{\chi}^{(3)} \colon \bigl(\bm{E}\otimes\bm{E}\otimes\bm{E}\bigr)
\;+\;\dots\,,
\end{equation}
where $\tens{\chi}^{(N)}$ is an $(N+1)$-rank tensor representing the $N$th-order nonlinear susceptibility, characterizing the material's ability to generate $N$th-order harmonic~\cite{boydNonlinearOptics2019}.

Generally, the components of the $(N+1)$th-order nonlinear susceptibility tensor 
$\chi^{(N+1)}_{i_1\cdots i_{N+1}}$ are typically 8–10 orders of magnitude smaller than those of 
$\chi^{(N)}_{i_1\cdots i_N}$. At modest field amplitudes, polarization remains essentially synchronous with the incident field, underpinning linear optical phenomena such as reflection and transmission. SHG, which involves the coherent combination of two photons at the fundamental frequency to produce a single photon at twice the original frequency, specifically arises from the second-order nonlinear polarization: $\bm{P}_{2\omega} = \tens{\chi}^{(2)}:\bm{E}_{\omega}\otimes\bm{E}_{\omega}$. Thus only can be observed on strong intensity incidence. Furthermore, considering the interaction between the linear polarization at the $2\omega$, $\bm{P}_{2\omega} = \tens{\chi}^{(1)}:\bm{E}_{2\omega}$, and the nonlinear polarization generated by the FW, $\bm{P}_{2\omega} = \tens{\chi}^{(1)}:\bm{E}_{\omega}\otimes\bm{E}_{\omega}$, the realization of detectable SHG signal requires appropriate matching conditions between the fundamental wave~(FW) and second harmonic wave~(SH), including phase matching of the propagating waves and mode matching of the standing wave (\textbf{Figure 2a}). In the structure with translational symmetry that support propagating mode with propagation constant, the phase matching: 
\begin{equation}
    \Delta \bm{k}=2\bm{k}(\omega) - \bm{k}(2\omega) = 0\,
\end{equation}
is required, where $\bm{k}(\omega) \text{ and }\bm{k}(2\omega)$ are wavevectors of the fundamental and second-harmonic waves, respectively. In contrast, structures with translational symmetry broken such as the unit cells of metasurfaces, conventional phase matching is no longer applicable due to the absence of propagating modes. Instead, resonant enhancement of the fundamental electric field enables the excitation of specific eigenmodes, which can strongly localize and amplify the nonlinear interaction within the nanostructure. Here, mode matching $\varkappa_{12}$, characterized by spatial and spectral overlap between fundamental and SH modes~\cite{2015NatureNanotechCelebranoModeMatchingMultiresonant,2020ScienceKoshelevSubwavelengthDielectricResonators,2019OpticaMinkovDoublyResonantH2}:
\begin{equation}
\varkappa_{12}
=
\underbrace{\frac{2\,\omega_{\mathrm{FW}}}
     {2\,\omega_{\mathrm{FW}} - \omega_{\mathrm{SH}} + i\,\gamma_{\mathrm{SH}}}}_{\text{Spectral Overlap}}\times
\underbrace{
\int_{V}
\bm{E}_{\mathrm{SH}}
\;\cdot\;
\bigl[
  \tens{\chi}^{(2)} \colon (\bm{E}_{\mathrm{FW}}\otimes\bm{E}_{\mathrm{FW}})
\bigr]
\;\mathrm{d}^3r}_{\text{Spatial Overlap}}\,,
\end{equation} 
becomes crucial~\cite{2018Adv.Opt.Photon.Keren-ZurShapingLightNonlinear,2019Adv.Photon.SainNonlinearOpticsAlldielectric}, as it is shown in~\textbf{Figure 2b}. One can see that the nonlinear conversion is governed by the spatial overlap integral between interacting modes, however, this overlap can vanish due to symmetry reasons for high-symmetry modes. Therefore, the selection rules for nonlinear conversion are determined by the combined influence of the resonator’s geometrical symmetry and the orientation of the nonlinear susceptibility tensor~\cite{frizyuk2019second}. For atomically thin dimension, such as two-dimensional materials, quantum dots, or quantum wells, neither phase nor mode matching is required, as quantum effects predominantly govern SHG~\cite{2022AdvFunctMaterialsKhanOpticalHarmonicGeneration2DMaterialsa}.

The nonlinear susceptibility $\tens{\chi}^{(2)}$ (with SI units $\text{m V}^{-1}$) originates from the asymmetry term of the electronic potential energy function~\cite{cazzanelliSecondOrderOptical2016} which is highly sensitive to material symmetry. In centrosymmetric materials, inversion symmetry implies the reverse SH radiation of the electron response at $\bm{k} $ and $-\bm{k}$ momentum. This identity reverse polarization results in the even symmetry of the electron potential in real space, leading to the counteract of second-harmonic contributions across the material, as depicted in \textbf{Figure 3a}. Formally, in mathematical terms, under inversion symmetry, 
\begin{equation}
\bm{P}_{2\omega}
\;=\;
\tens{\chi}^{(2)}\bigl(-\bm{r}\bigr)
\colon
\bigl[
  \bigl(-\bm{E}_{\omega}\bigr)
  \otimes
  \bigl(-\bm{E}_{\omega}\bigr)
\bigr]
\;=\;
-\bm{P}_{2\omega}\,.
\end{equation}
necessitating $\tens{\chi}^{(2)}(-\bm{r}) = -\tens{\chi}^{(2)}(\bm{r})$, implying $\tens{\chi}^{(2)} = 0$. Thus, natural SHG inherently occurs only in noncentrosymmetric materials~\cite{2022OEAGigliAlldielectricH2Metasurfaces}. However, applying a electrostatic field $\bm{E}_\text{DC}$ in centrosymmetric materials distorts the electron cloud, breaking the inversion symmetry, and producing an effective second-order susceptibility via the interaction of a third-order nonlinear susceptibility with electrostatic field. Therefore, the SHG response at momentum states $\pm \bm{k}$ exhibits a disparity and the potential energy becomes non-even (\textbf{Figures 3c} and  \textbf{3e}).

It is important to clearly distinguish EFISH from similar effects such as CISH~\cite{2012Phys.Rev.Lett.RuzickaSecondHarmonicGenerationInduced} and QCSE~\cite{2022Nat.Photon.YuElectricallyTunableNonlinear}, all of which can electrically modulate the $\tens{\chi}^{(2)}_{\text{eff}}$. While CISH also breaks centrosymmetry through current bias, however, the underlying symmetry-breaking mechanism is fundamentally different from EFISH. 

In detail, EFISH originates from the asymmetry electron potential induced by a space-charge field. However, in the materials with high carriers density, the internal electrostatic field is effectively screened, rendering EFISH ineffective. In such cases, symmetry breaking is instead driven by an applied electrical current, named CISH. It is the oblique fermi level that distorts the symmetry of the electron distribution in the momentum space (\textbf{Figure~3d}), leading to incomplete cancellation of the SHG with wavevectors $\bm{k}$ and $-\bm{k}$. 

Another phenomenon, QCSE, predominates in low-dimensional systems such as vdW materials and quantum engineered heterostructures. In these systems, quantum confinement gives rise to excitons, quasiparticles comprising Coulomb-bound electron–hole pairs, and intersubband transitions (ISTs). Both excitonic and IST features appear as sharp resonances in the second-order nonlinear susceptibility spectrum, governed by the electronic bandgap~\cite{1983IEEEJ.QuantumElectron.GurnickSyntheticNonlinearSemiconductors}. Electrical modulation via QCSE shifts the electronic band structure and thereby tunes these resonances, modifying the nonlinear optical response~(\textbf{Figure 3f, g}).

Generally, EFISH can be expressed as~\cite{1991Phys.Rev.ABavliRelationshipSecondharmonicGeneration}:
\begin{equation}
\mathbf{P}_{2\omega} = (\tens{\chi}^{(2)} + \underbrace{3\tens{\chi}^{(3)} : \bm{E}_{\text{DC}}}_{\tens{\chi}^{(2)}_\text{eff}\propto\bm{E}_{\text{DC}}}) : \bm{E}_{\omega}\otimes\bm{E}_{\omega}\,,\label{eq:P2omega}
\end{equation}
where $\tens{\chi}^{(2)}$ is the intrinsic nonlinear susceptibility, and $3\tens{\chi}^{(3)}:\bm{E}_{\text{DC}}$ represents the effective second-order susceptibility induced by the EFISH effect. In noncentrosymmetric materials,however, $\tens{\chi}^{(3)}$ is typically much smaller than $\tens{\chi}^{(2)}$, rendering EFISH negligible compared to intrinsic SHG. Hence, EFISH studies predominantly focus on centrosymmetric materials ($\tens{\chi}^{(2)} = 0$), where all SH signals arise from EFISH, enabling significant modulation depths. Thus, EFISH research primarily addresses three key aspects derived from~\cref{eq:P2omega}: (i) $\tens{\chi}^{(3)}$: selecting nonlinear optical materials, (ii) $\bm{E}_\text{DC}$: engineering static electric field, and (iii) $\bm{E}_{\omega}$: manipulating fundamental electric field distributions. Specifically, while aspects (i) and (ii) involve material properties, (i) emphasizes intrinsic nonlinear optical characteristics, whereas (ii) highlights semiconductor electrical properties. Moreover, aspects (ii) and (iii) involve structural considerations: aspect (ii) addresses electrical structure designs (e.g., metal–semiconductor–metal or p–i–n junctions), whereas aspect (iii) emphasizes photonics structure engineering, such as metasurface, waveguide and micro cavity.

\subsection{Selection of nonlinear optical materials}

\subsubsection{Bulk crystals}

Within crystalline materials, crystal structures can be classified into 32 crystal classes, of which 11 distinct crystallographic point groups possess centrosymmetry~\cite{kittel2018introduction}. For example, calcite, which is the first material discovered to exist EFISH signal by Terhune et~al.~\cite{1962Phys.Rev.Lett.TerhuneOpticalHarmonicGeneration}, belongs to the symmetry group $D_{3d}$. The measured SHG signal exhibits a parabolic dependence on the applied external voltage, providing unambiguous evidence for the EFISH relationship: $\tens{\chi}^{(2)}_{\text{eff}} = 3\tens{\chi}^{(3)}:\bm{E}_{\text{DC}}$. 

In addition, there are materials exhibiting unique electron polarization behavior such as ferroelectric systems~\cite{2005Rev.Mod.Phys.DawberPhysicsThinfilmFerroelectric}, donor-accptor systems~\cite{wan2020acceptor}, and the topological state~\cite{vanderbilt2018berry},which display peculiar EFISH response. Ferroelectric materials, which exhibit unique electronic characteristics arising from interactions leading to a double-well potential with two or more stable polarization states in the absence of an electric field~\cite{2005Rev.Mod.Phys.DawberPhysicsThinfilmFerroelectric,2016NatRevMaterMartinThinfilmFerroelectricMaterials,2018AdvancedMaterialsZhangFerroelectricPiezoelectricEffects,2020NatElectronKhanFutureFerroelectricFieldeffect}, have garnered significant attention in electrically tunable nonlinear photonics. These polarization states can be switched by applying an external electric field, producing a characteristic hysteresis loop due to the hysteretic relationship between the electric displacement and applied field~\cite{1998NatureFiorePhaseMatchingUsing}. Correspondingly, the linear dielectric polarization displays a butterfly-shaped loop as a function of the applied field~\cite{2005NatureMaterKingonLeadZirconateTitanate,2019AdvFunctMaterialsSilvaHighPerformanceFerroelectricDielectric}. Furthermore, these materials are inherently noncentrosymmetric~\cite{2011JournaloftheAmericanCeramicSocietyDenevProbingFerroelectricsUsing}, thus exhibiting high intrinsic second-order nonlinear susceptibility, $\tens{\chi}^{(2)}$~\cite{2022Nat.Photon.AbdelwahabGiantSecondharmonicGenerationa}. Consequently, their second-order dielectric polarization is influenced by both intrinsic ferroelectric properties and the EFISH effect. Gilles et~al.~\cite{2021AdvancedOpticalMaterialsFeutmbaReversibleTunableSecondOrder} reported the electrically tunable SHG in the lead zirconate titanate (PZT) thin film via both ferroelectric property and EFISH (\textbf{Figure~4a}). The maximum value of $\chi^{(2)}_{zzz} = 128\,\text{pm V}^{-1}$ is measured at a bias of $-20\,\text{V}$ with a modulation depth of $3.35\,\text{pm}^2\,\text{V}^{-2}$. A butterfly-shaped hysteresis behavior in SH power as a function of DC bias confirms the ferroelectric nature of the PZT thin film (\textbf{Figure~4b}). It is noteworthy that the relationship between SH power and applied field differs in small versus large voltage regimes: a parabolic dependence at small absolute fields indicative of EFISH, and linear behavior at large absolute fields attributed to changes in ferroelectric domain configuration.

In addition to uniform spontaneous polarization, ferroelectric systems exhibit polar skyrmion, which is a non-trivial, topologically protected configuration of polarization, analogous to vortex-like domain walls possessing centrosymmetric characteristics~\cite{2024AdvFunctMaterialsXuTunableFerroelectricTopological,2023Rev.Mod.Phys.JunqueraTopologicalPhasesPolar,2021AdvancedMaterialsChenRecentProgressTopological,2022NatureHanHighdensitySwitchableSkyrmionlike}. This phenomenon have been discovered in many system like $\text{PbTiO}_3/\text{SrTiO}_3$ superlattices~\cite{2022NatureRusuFerroelectricIncommensurateSpin,2020Nat.Mater.WangPolarMeronLattice,2019NatureDasObservationRoomtemperaturePolar,2017NatCommunLiQuantificationFlexoelectricityPbTiO3,2016NatureYadavObservationPolarVortices,2015ScienceTangObservationPeriodicArray}, where the polarization forms a nanoscale-size periodic unit with swirling electric dipoles and long-range in-plane ordering. In situ studies of the polarization evolution in these structures reveal their strong responsiveness to external stimuli, resulting in significant alterations in the magnitude of polarization. These findings highlight the potential of topological polar structures for EFISH with giant modulation depth. Recently, Sixu et~al.~\cite{2024NatCommunWangGiantElectricFieldinduceda} observed a huge SHG response (${\chi}^{(2)}_\text{eff,zzz} = -54.2 \, \text{pm V}^{-1}$) in polar skyrmions through the EFISH effect in \\ 
$\text{PbTiO}_3/\text{SrTiO}_3$ superlattice with -14~V bias (\textbf{Figure~4c, e}), thereby expanding the range of material property for EFISH. The three-dimensional X-ray reciprocal space mapping (RSM) around the (001)-oriented $\left(\mathrm{LaAlO}_3\right)_{0.3}\left(\mathrm{SrAl}_{0.5}\mathrm{Ta}_{0.5}\mathrm{O}_3\right)_{0.7}$ (LSAT) substrate reflection, accompanied by two in-plane RSMs, demonstrates excellent crystallinity of the fabricated superlattices (\textbf{Figure~4d}). Along the in-plane directions, the observed fourfold lobe pattern of satellite peaks indicates the formation of abundant polar skyrmions, exhibiting significant in-plane ordering aligned along the [100] and [010] crystallographic directions (\textbf{Figure~4d}, inset). In this ground state, skyrmions are $\tens{\chi}^{(2)}$ inactive due to the pseudo centrosymmetry of the dipoles within each skyrmion and the surrounding interfacial $c$-domain region. This symmetry can be dramatically broken by the application of an electric bias, and the significant alterations in the magnitude of polarization enabling a high-modulation depth EFISH of $-664\%\,\text{V}^{-1}$.

\subsubsection{Van der Waals materials}

Van der Waals layered materials are crystal consisting of crystalline sheets held together by strong in-plane covalent bonds and weak out-of-plane vdW interactions~\cite{2022AdvFunctMaterialsKhanOpticalHarmonicGeneration2DMaterialsa,Adv.Photon.2024WangTwodimensionalMaterialsTunable,huangSecondHarmonicGeneration2024}. This weak interlayer force facilitates the mechanical exfoliation of the materials into two-dimensional (2D) monolayers and few-layer crystals. The symmetry of few-layer 2D materials varies with the number of layers~\cite{2013NanoLett.LiProbingSymmetryProperties}, influencing the nonlinear susceptibility tensor, characterized by angular-dependent nonlinear optical signals. Moreover, layered 2D materials with arbitrary interlayer twist angles~\cite{2021Sci.Adv.YaoEnhancedTunableSecond,2019NatureAlexeevResonantlyHybridizedExcitonsa}, wherein layers remain bound by vdW interactions while preserving their individual crystal symmetries and forming moiré superlattices, have also been realized. Twisted stacking of 2DLMs with moiré superlattices features twist angle-dependent crystal symmetry and enables the exploration of special moiré excitons resonances in artificial twisted homotrilayers and heterobilayers in artificial twisted homo- and heterostructures. The resulting symmetry variations in arbitrarily stacking order~\cite{2022AdvFunctMaterialsKhanOpticalHarmonicGeneration2DMaterialsa,2022AdvancedOpticalMaterialsUllahHarmonicGenerationLowDimensional} and angle open unprecedented opportunities for SHG, including the EFISH effect in inversion-symmetric layered 2D materials. 

However, unlike EFISH in bulk crystals,where symmetry breaking can be understood simply as an electrostatic field bias influencing the electron potential energy (macroscopically described by the tensor product of $\tens{\chi}^{(3)}$ and $\bm{E}_{\text{DC}}$ with a constant $\tens{\chi}^{(3)}$), electron behaviors in vdW materials under electrostatic fields are significantly more complex. Such behaviors are strongly influenced by rich quantum effects, leading to different electron behaviors in different layers (modifications in the $\tens{\chi}^{(3)}$ tensor itself). Consequently, descriptions of EFISH in vdW materials become intricately linked to these quantum phenomena. For example, various multilayered 2D materials exhibit centrosymmetric properties~\cite{2013NanoLett.LiProbingSymmetryProperties}, such as bilayer graphene (BLG)~\cite{2018Sci.Adv.ShanStackingSymmetryGoverned}, bilayer 2H transition metal dichalcogenides (TMDs) with M = Mo, W and X = S, Se)~\cite{2021Adv.Mater.WangStackingEngineeredHeterostructuresTransition}, and trilayer ReS\textsubscript{2}. The different stacking methods significantly influence their symmetry~\cite{geim2013van,doi:10.1126/science.aac9439}: 2H and AB stacking patterns belong to the centrosymmetric $D_{3d}$ point group, while AA and 1T polymorphs belong to the centrosymmetric $D_{3h}$ group. Specifically, the $D_{3d}$ group lacks the out-of-plane third-order susceptibility elements required for EFISH under vertical electrostatic bias. Microscopically, the absence of EFISH in these structures arises from oppositely oriented in-plane dipole moments induced by a uniform electrostatic field in each layer, preserving overall centrosymmetry. Nonetheless, phenomena have been observed wherein electrostatic bias induces symmetry breaking due to differential perturbations between upper and lower layers~\cite{2015Phys.Rev.BBrunIntenseTunableSecondharmonic}. Therefore, this review collectively addresses various centrosymmetry-breaking mechanisms responsible for SHG in centrosymmetric vdW materials, including charge-induced SHG, exciton charging, and electrostatic doping.

For example, monolayer MoTe\textsubscript{2} exhibits intriguing phase-dependent symmetry characteristics, being centrosymmetric in the 1T' phase and noncentrosymmetric in the 1H phase. Wang et~al. demonstrated that these distinct symmetry phases of MoTe\textsubscript{2} can be electrically controlled via electrostatic doping, observing a pronounced decrease in broadband SHG intensity associated with the transition from the noncentrosymmetric to centrosymmetric phase at a critical gate bias (approximately 3~V)~\cite{2021NatElectronWangDirectElectricalModulation}, as illustrated in \textbf{Figure~5a}. This electrostatically driven phase modulation enables highly efficient SHG switching, characterized by an exceptional on/off ratio of 1,000, a modulation strength of approximately 30,000\% per volt, and a broad operational spectral range of around 300~nm. In particular, such phase transitions not only facilitate symmetry control in monolayers but also enable symmetry breaking in bilayer 2H-MoTe\textsubscript{2}. It is only the upper layer undergoes the phase transition under electrostatic doping, while the lower layer retains its original phase that result in a distinct SHG response different from that observed in monolayer MoTe\textsubscript{2} (\textbf{Figure~5b}).

Similarly, Huakang et~al.~\cite{2015NanoLett.YuChargeInducedSecondHarmonicGeneration} observed pronounced SHG in 2H-stacked bilayer WSe\textsubscript{2}, induced by a back-gate voltage below the saturation point (approximately $-20\,\text{V}$), as illustrated in \textbf{Figure~5c}. This phenomenon cannot be explained by conventional EFISH models, as bilayer 2H-WSe\textsubscript{2}, belonging to the $D_{3d}$ point group, lacks the necessary out-of-plane $\tens{\chi}^{(3)}$ component for EFISH. Microscopically, the second-order bond hyperpolarizabilities induced by a uniform electric field in the upper and lower layers have opposite orientations, leading to cancellation and thus no net EFISH. However, a negative gate voltage induces hole accumulation predominantly in the bottom WSe\textsubscript{2} layer, specifically forming localized W $5d_{x^2 - y^2, xy}$ orbitals, which prevents further electric field penetration into the upper layer. This asymmetric charge distribution breaks the inversion symmetry, preventing cancellation of second-order bond hyperpolarizabilities and resulting in observable SHG. Under positive voltage bias, in contrast, only a depletion regime forms without significant charge accumulation; consequently, symmetry breaking is minimal, and the resulting SHG signal remains negligible. This phenomenon, termed charge-induced SHG, arises explicitly from the layer-dependent spatial distribution of accumulated charges. A similar behavior was also observed in trilayer ReS\textsubscript{2}~\cite{2022ACSNanoWangElectricallyTunableSecond}, as illustrated in \textbf{Figure~5d}. However, in contrast to the explanation for hole accumulation for bilayer WSe\textsubscript{2}, Wang et~al. attributed the observed SHG to interlayer charge transfer, causing a vertical asymmetry in charge density distribution as supported by first-principles calculations. In trilayer ReS\textsubscript{2}, a weak SHG signal under positive gate voltage arises due to electron accumulation at the bottom sulfur layer, which leads to charge shielding. Interestingly, a decrease in SHG signal at gate voltages above $20\,\text{V}$ remains unexplained, highlighting the need for further exploration of complex interlayer charge dynamics in vdW multilayers.

In addition to symmetry breaking induced by perturbations in across layers electrons, exciton resonances in TMDs within the visible and near-infrared spectral regime (approximately $1.5$–$2\,\text{eV}$) significantly enhance the light–matter interaction. Differences exciton behavior between layers can dramatically amplify SHG signals. For example, Soonyoung et~al.~\cite{2024NanoLett.ChaEnhancingResonantSecondHarmonic} reported a 40-fold SHG enhancement in bilayer WeS\textsubscript{2} by leveraging strong exciton resonances and a layer-dependent exciton–polaron effect. By selectively localizing injected holes within one layer, exciton–polaron states are induced in the hole-rich layer, while normal exciton states persist in the charge-neutral layer. This distinct resonance condition effectively breaks interlayer inversion symmetry, thus promoting resonant SHG (\textbf{Figure~5e}). 
Moreover, symmetry breaking induced by a perpendicular electrostatic field in bilayer MoS\textsubscript{2}~\cite{2013NanoLett.LiProbingSymmetryProperties}, distinct exciton characteristics within each layer due to hybridization of sulfur orbitals, results in a 60-fold enhancement of the EFISH signal due to interlayer coupling at an SHG energy of $E_{2\omega} = 2.49~\text{eV}$ (\textbf{Figure~5f}) ~\cite{2017NanoLett.KleinElectricFieldSwitchableSecondHarmonic}. This symmetry breaking can be visualized via calculations of the two-particle exciton wave functions over the Brillouin zone (BZ) (as illustrated in Figure~6 of the original work~\cite{2017NanoLett.KleinElectricFieldSwitchableSecondHarmonic}. At C resonance, characteristic characteristics of the wave function appear between the $\Gamma$ and M points, exhibiting minimal modification under external electric fields. In contrast, at a slight detuning of, the wave function show pronounced features near the $\Gamma$ point.) Due to differential contributions from atomic orbitals to Bloch states in this region of the BZ, the exciton wavefunctions exhibit significant symmetry breaking below the C-resonance. Recently, Daichi Okada et~al.\cite{2025J.Am.Chem.Soc.OkadaElectricFieldInducedGiantResonant} publish that the exciton resonance enhancement EFISH also exist in Two-Dimensional Hybrid Perovskite. The authors employed $(\mathrm{BA})_2(\mathrm{MA})_{n-1}\mathrm{Pb}_n\mathrm{I}_{3n+1}$ (with $n = 2$) as the host material and introduced a polar organic cation, R-MBACl, into the perovskite lattice. Under the applied vertical field, the polar R-MBA$^+$ molecules reorient along the field direction, locally breaking centrosymmetry and enabling the EFISH process. Remarkably, when the SHG photon energy is tuned near the exciton resonance, the system exhibits a resonantly enhanced nonlinear response, with SHG intensity boosted by nearly two orders of magnitude.

Other two-dimensional materials, such as group IIIA–VIA compounds (e.g., GaSe, InSe, In\textsubscript{2}Se$_3$, Ga\textsubscript{2}Se$_3$), group VIA monochalcogenides (MX, e.g., GeS, GeSe, SnS and SnSe) and hexagonal boron nitride (h-BN), have also shown great potential in nonlinear optics~\cite{2017NanoLett.WangGiantOpticalSeconda}. In particular, GaSe and InSe only exist in three distinct crystalline structure types ($\epsilon$, $\beta$, and $\gamma$) with the even-layered $\beta$-type stacking exhibiting centrosymmetry~\cite{2014AdvancedMaterialsFengBackGatedMultilayer,2015AdvancedOpticalMaterialsHoBendingPhotoluminescenceSurface,2015AngewChemIntEdGhoshDesigningElasticOrganic,2019NanoLett.HaoPhaseIdentificationStrong}. However, the EFISH effect has not yet been investigated in this material, which is desirable to expect. Furthermore, MX monochalcogenides have been predicted to be two-dimensional ferroelectric materials~\cite{2016NanoLett.WuIntrinsicFerroelasticityMultiferroicity,2019AdvancedMaterialsChangEnhancedSpontaneousPolarization}. Inspired by the EFISH effect in bulk ferroelectric materials~\cite{2021AdvancedOpticalMaterialsFeutmbaReversibleTunableSecondOrder,2024NatCommunWangGiantElectricFieldinduceda}, it is anticipated that EFISH in group VIA monochalcogenides will exhibit outstanding performance in the future. 

\subsubsection{Polymers}

In addition to inorganic crystals, organic conjugated polymer, particularly $\pi$-conjugated systems, facilitating significant electron delocalization, which enhances both the linear polarizability and higher-order hyperpolarizabilities~\cite{1986Appl.Phys.Lett.SingerSecondHarmonicGeneration,2000J.Am.Chem.Soc.ClarkSecondHarmonicGeneration,2000Chem.Rev.DelaireLinearNonlinearOptical,2008ProgressinPolymerScienceChoRecentProgressSecondorder}, making them promising materials for optoelectronic applications~\cite{2004Opt.Lett.JuSecondharmonicGenerationPeriodically}. Their nonlinear response arises from the delocalized $\pi$-electron system, which enables efficient light-matter interaction, charge transport, and modulation of optical properties under external stimuli~\cite{2008NatureMaterYapSimultaneousOptimizationChargecarrier}, showing promising potential in EFISH. Shumei et~al.~\cite{2019LightSciApplChenGiganticElectricfieldinducedSecond} reported that substantial EFISH enhancement can be achieved due to band-edge effects in the organic conjugated polymer (PFO, poly(9,9-di-$n$-dodecylfluorenyl-2,7-diyl)). Their device consists of a 100~nm thick thin film of the organic conjugated polymer PFO sandwiched between aluminum and ITO, as illustrated in \textbf{Figure 4f}. With an incident fundamental wave having a 45-degree transverse magnetic (TM) polarization, the reflection spectrum exhibits strong absorption below 400~nm (\textbf{Figure 4g}), showing the bandgap energy of ~2.95 eV. When the fundamental photon energy is half of the peak absorption energy (840~nm wavelength), a 25-fold enhancement in SHG intensity (under 6~V bias) with a modulation ratio of up to 422\%~V$^{-1}$.

\subsection{Engineering of $\bm{E}_\text{DC}$}

The static electrical field arises from spatially non-uniform free-charge distributions and is described by Poisson's equation~\cite{kong1975theory}. Common charged species include free electrons and holes, lattice ions, ionized dopants or defects, and charged quasiparticles such as polarons and Cooper pairs~\cite{devreese2013electronic}. Engineering $\bm{E}_\text{DC}$ for EFISH therefore relies on the manipulation of these charged species.

\subsubsection{External electrodes}

The most established method to introduce a static electric field is the application of external electrodes manipulating free carriers. Due to their abundant and delocalized conduction electrons, metals instantaneously screen internal electric fields when a potential is applied, thereby confining excess charges to their surfaces and maintaining an equipotential state. The surface charges formed in this process generate an external electrostatic field with a spatial profile determined by the electrode geometry and applied bias. Such precise manipulation of boundary potentials and electric field distributions enables effective modulation of static charge distributions in adjacent materials.

When an insulating material is placed between external electrodes, the static electric field distribution is predominantly determined by the geometry and potentials of the electrodes, with the neglected influence of free carriers or internal charges in the insulating material. Consequently, excellent agreement between the simulated static electric field distribution and the measured EFISH response is observed. For example, calcite\cite{1962Phys.Rev.Lett.TerhuneOpticalHarmonicGeneration} and silicon oxide\cite{2022Opt.ExpressWidhalmElectricfieldinducedSecondHarmonic}. In their work, the assumption of a homogeneous internal field successfully explained the observed quartic dependence of the second-harmonic generation intensity on the applied voltage.

However, in semiconductor devices featuring metal oxide semiconductor (MOS)~\cite{2022Opt.ExpressWidhalmElectricfieldinducedSecondHarmonic} or metal semiconductor (MSM)~\cite{2018NatCommunRenStrongModulationSecondharmonic} structures, carrier dynamics in the semiconductor layer under bias cannot be neglected. At thermal equilibrium, the alignment of Fermi levels at the metal (or metal-oxide)--semiconductor interface results in the formation of a Schottky barrier. This barrier induces carrier depletion and establishes a built-in electrostatic field within the depletion region. Conversely, the electrically neutral regions with high free-carrier densities, effective screening suppresses internal electrostatic fields~\cite{boer2010introduction}. Consequently, EFISH signals primarily arise within depletion regions. Application of an external voltage modulates both the width and magnitude of the depletion-region field, leading to a linear relationship between the electrostatic field and applied bias. Experimentally, this field modulation manifests as a parabolic dependence of the EFISH signal intensity on the applied voltage within the depletion regime~\cite{1994Opt.Lett.AktsipetrovOpticalSecondharmonicGeneration,1996Phys.Rev.BAktsipetrovDcelectricfieldinducedSecondharmonicGeneration}.

Beyond the classical Schottky mechanism, certain materials exhibit high-field domain formation capable of sustaining strong and uniform internal electrostatic fields. For example, in GaAs~\cite{gunn1963microwave}, high-field domains emerge due to electron scattering into satellite valleys of the conduction band, thus reducing electron mobility at elevated fields. Similarly, in CdS~\cite{boer1965layer,boer1968stationary}, holes trapped at the defect sites become ionized under strong fields and subsequently recombine with conduction electrons, decreasing the free carrier concentration. This field-quenching mechanism results in the establishment of distinct high- and low-field domains. Recently, high-field domains have been experimentally observed in CdS nanostructures~\cite{2018NatCommunRenStrongModulationSecondharmonic}, as depicted in \textbf{Figure~6a}. As the reverse bias voltage increases, the Schottky barrier broadens, triggering the formation of high-field domains and consequently yielding nonlinear SHG intensity voltage characteristics. Three distinct operational regimes are discernible (\textbf{Figure~6b}): (i) weak EFISH dominated by the Schottky field for $V_\text{DS} < V_\text{T}$; (ii) formation and growth of a high-field domain for $V_\text{T} < V_\text{DS} < V_\text{C}$; and (iii) saturation of field amplitude for $V_\text{DS} > V_\text{C}$. The high-field domain formation originates from electron excitation from the valence band to deep acceptor states, which upon ionization become negatively charged, stabilizing the domain. Increasing the voltage further expands the spatial extent of this domain until saturation occurs.

Another effective way to establish a nearly uniform internal electric field is the p–i–n junction. In this architecture, carrier diffusion across the p–i and i–n interfaces creates a wide depletion region in the intrinsic layer; the uncompensated ionized dopants in that region then sustain a built-in electric field throughout the i-region. Timurdogan et~al.~\cite{2017NaturePhotonTimurdoganElectricFieldinducedSecondorder} design an EFISH-enabled quasi-phase matching (QPM) grating in silicon p–i–n waveguides (\textbf{Figure~6c}). By engineering periodic p–i–n junctions, quasi-phase matching between fundamental and second-harmonic is achieved, attaining a peak conversion efficiency of $13~\text{W}^{-1}$ and an effective second-order susceptibility $\chi^{(2)}_{\text{eff}} = 41\,\text{pm V}^{-1}$ at a wavelength of $2.29\,\mu\text{m}$ (\textbf{Figure~6d}).

In addition, Jasinskas et~al.~\cite{2016Opt.Lett.JasinskasBackgroundfreeElectricFieldinduced} report a EFISH design using inter-digitated electrode structures. They employed direct laser ablation to pattern comb‐shaped electrodes in a thin chromium film deposited on a transparent glass substrate. An organic polymer layer was then coated atop these electrodes as the NLO material~(Figure~6e). By applying alternating DC biases across adjacent electrode fingers, a spatially periodic nonlinear‐susceptibility grating of alternating $+\tens{\chi}^{(2)}_{\text{eff}}$ and $-\tens{\chi}^{(2)}_{\text{eff}}$ regions is established, enabling second‐harmonic diffraction (\textbf{Figure~6f}).

Apart from external DC voltage biasing, internal DC fields can also be induced by the Maxwell–Wagner (MW) effect under high-frequency AC voltages~\cite{wagner1914erklarung}. The MW effect originates from charge accumulation at interfaces between materials with mismatched dielectric constants and conductivities. Although MW-induced EFISH (MW-EFISH) is generally considered negligible due to weak and localized field formation at sharp interfaces, graded interfaces formed via diffusion can enable more extended internal DC fields. Recently, Scherbak et~al.~\cite{2024JAmCeramSoc.ScherbakMaxwellWagnerEffecta} demonstrated MW-EFISH in ion-exchanged soda-lime glass, where accumulated interfacial charge generated internal fields as high as $10^8\,\text{V/ m}$, boosting SHG by over two orders of magnitude.

\subsubsection{Photo-induced field}
In addition to externally applied biases, another viable approach to generate a static electric field involves optically induced spatial carrier redistribution through asymmetric carrier trapping by diffusion~\cite{1996Phys.Rev.Lett.BlochElectronPhotoinjectionSilicon,1995Opt.Lett.MihaychukTimedependentSecondharmonicGeneration} or electron emission~\cite{1986Opt.Lett.OsterbergDyeLaserPumped,1995QuantumElectron.DianovPhotoinducedGenerationSecond,1996Phys.Rev.Lett.AtanasovCoherentControlPhotocurrenta}, giving rise to an all-optical EFISH effect. In this scenario, the photon induced spatial accumulation of separated electrons and holes generates a static field; however, this built-in field decays over time due to carrier recombination. The lifetime of these optically induced static fields varies considerably, ranging from femtoseconds in ultrafast processes to seconds or even longer in defect-stabilized systems, largely dependent on the depth of the acceptor states. Consequently, the induced field typically increases over the first few hundred seconds of illumination, gradually saturates, and then decays once the illumination is turned off.

Generally, it is multiple-photon absorption processes transfer free electrons from donor into acceptor, simultaneously leaving behind holes that generate the electrostatic field on the interface. This process is governed by three critical factors: (i) density of states (DOS)—high DOS in the electron donor and acceptor enhances free electron excitation probability and electron injection likelihood respectively; (ii) band alignment—the energy level difference between the donor conduction band and the acceptor conduction band, determining the injection barrier; and (iii) carrier lifetime—the lifetime of trapped electrons directly determines the longevity of the electrostatic field.

Charge‐transfer–induced EFISH has been extensively studied at Si–SiO$_2$ interfaces, where silicon and SiO$_2$ acts as the electron donor and acceptor respectively ~\cite{1998PHCREVEWETTERSWangCoupledElectronHoleDynamics,2002J.Appl.Phys.FomenkoSecondHarmonicGeneration,2002Phys.Rev.BGlinkaCharacterizationChargecarrierDynamics,2002J.Phys.Chem.BMitchellChargeTrappingChemically,2003Phys.Rev.BFomenkoCombinedElectronholeDynamics,2004Phys.Rev.BScheidtChargecarrierDynamicsTrap,2008J.Appl.Phys.ScheidtIonizationShieldingInterface,2011Appl.Phys.Lett.FioreSecondHarmonicGeneration}. The silicon band gap of approximately $1.1\,\mathrm{eV}$ and the offset of the conduction band $\sim3.2\,\mathrm{eV}$ to SiO$_2$ require multiphoton energy exceed $\sim4.3\,\mathrm{eV}$. When the fundamental photon energy exceeds $1.36\,\mathrm{eV}$, electrons in silicon can be promoted into defect states at the SiO$_2$–Si interface via three‐photon absorption or direct absorption of the third‐harmonic photon~\cite{1996Phys.Rev.Lett.BlochElectronPhotoinjectionSilicon}. At higher photon energies and excitation intensities, increased free‐carrier generation and associated space‐charge screening lead to a deviation from quadratic scaling in the SHG response. Once the field saturates, the SHG intensity scales quadratically with pump power, reflecting an intensity‐independent steady‐state field~\cite{2001Phys.Rev.BFomenkoNonquadraticSecondharmonicGeneration,1995Opt.Lett.MihaychukTimedependentSecondharmonicGeneration}.
In particular, a SHG enhancement can be observed after brief interruptions of illumination have been attributed to four-photon absorption–induced hole injection into the SiO$_2$ valence band when the photon energy exceeds $1.52\,\mathrm{eV}$. This mechanism arises because the low mobility of injected holes prolongs their accumulation at the semiconductor–oxide interface, thereby augmenting the built‐in electric field~\cite{1998PHCREVEWETTERSWangCoupledElectronHoleDynamics}.  

Except for the Si–SiO$_2$ bulk interfaces, Okada et~al.~\cite{2021AdvFunctMaterialsZhangAnomalouslyStrongSecondHarmonic} reported pronounced EFISH generation in quasi-centrosymmetric GaAs nanowire systems. Although bulk GaAs is intrinsically noncentrosymmetric, nanowires grown along certain crystallographic orientations (e.g., [111]) behave effectively centrosymmetric. The SHG enhancement was ascribed to diffusion of photon induced carriers and the ensuing EFISH within the nanowire. To verify this mechanism, the authors performed polarization-resolved, photo-modulation SHG measurements using a secondary 405 nm laser (above the GaAs bandgap) to generate free carriers that partially screen the internal electric field driving EFISH. Comparison of SHG intensities with and without the modulation beam revealed a pronounced suppression of the SHG signal, confirming the pivotal role of the built-in field. Notably, however, the SHG intensity exhibited a quadratic dependence on pump intensity, deviating from the non-quadratic dependence predicted by conventional all optical EFISH theory, suggesting further discussion should be considered.

Metal–semiconductor interfaces similarly exploit optically driven charge separation to sustain static fields. Unlike semiconductor–insulator interfaces—where only carriers (e.g., electrons) inject into the insulator, metal–semiconductor junctions allow both the metal and the semiconductor to function as electron donors or acceptors, depending on their work‐function difference. In particular, plasmonic mental nanostructures leverage LSPR to enhance near‐field intensities and facilitate ultrafast charge transfer~\cite{2013NatureNanotechGiugniHotelectronNanoscopyUsing}. For example, Yali et~al.~\cite{2023LightSci.Appl.SunAllopticalGenerationStatic} demonstrated a hybrid SHG nanoantenna comprising a gold nanoparticle atop a silicon “nano‐table” (\textbf{Figure~7a}). Under $1047\,\mathrm{nm}$ ($1.18\,\mathrm{eV}$) excitation, electrons transfer from silicon to gold, generating a static field of $\sim10^{8}\,\text{V/m}$ at the Si–Au interface. The sub‐quadratic scaling of SHG intensity with pump power in a log–log plot confirms the EFISH mechanism (\textbf{Figure~7b}), and the tens‐of‐seconds rise time to saturation reflects carrier trapping and detrapping at interface defects. Moreover, LSPR decay generates “hot” electrons with energies exceeding the Schottky barrier, which inject into adjacent semiconductor conduction bands on picosecond timescales, transiently breaking inversion symmetry and driving ultrafast EFISH~\cite{2020Nat.Mater.TagliabueUltrafastHotholeInjection,2009Phys.Rev.Lett.DevizisUltrafastDynamicsCarrier,2014NaturePhotonClaveroPlasmoninducedHotelectronGeneration,2017ACSPhotonicsBesteiroUnderstandingHotElectronGeneration,2020J.Appl.Phys.MascarettiHotElectronThermal,2024eLightKhurginHotelectronDynamicsPlasmonic}. Coulomb forces then return carriers to the metal over femto‐ to picosecond timescales \cite{2014NatCommunSundararamanTheoreticalPredictionsHotcarrier,2015NatureNanotechBrongersmaPlasmoninducedHotCarrier,2015NatCommunZhengDistinguishingPlasmoninducedPhotoexcited}. Wen et~al.~\cite{2018NanoLett.WenPlasmonicHotCarriersControlled} confirmed this mechanism by observing enhanced SHG from gold nanobars on bilayer $\mathrm{WSe_2}$  (\textbf{Figure~7c}), which was suppressed by inserting an hBN interlayer to block hot‐electron transfer. Similarly, Taghinejad et~al.~\cite{2020Phys.Rev.Lett.TaghinejadTransientSecondOrderNonlinearb} reported transient SHG in amorphous $\mathrm{TiO_2}$ induced by asymmetric hot‐carrier injection from triangular gold nanostructures, yielding SHG transients that mirror hot‐electron relaxation dynamics  (\textbf{Figure~7d}).

In addition to independent multiphoton absorption or plasmon-decay-induced free-electron generation, processes primarily governed by photon energy, an interesting phenomenon arises when considering the interference of FW and SH photon absorption processes named coherent photogalvanic effect~(CPE)~\cite{2017NatCommunBillatLargeSecondHarmonic,2019Nat.PhotonicsHicksteinSelforganizedNonlinearGratings,2021Nat.PhotonicsLuEfficientPhotoinducedSecondharmonic,2022Nat.Photon.NitissOpticallyReconfigurableQuasiphasematching}. This interaction induces asymmetric photoelectron emission. The subsequent capture of these asymmetrically emitted electrons by deep trap states at the boundaries of the illuminated region results in a substantial built-in electrostatic field. The total ionization rate from multiphoton absorption can be described by~\cite{1991Opt.Lett.AndersonModelSecondharmonicGeneration}:
\begin{equation}
\begin{aligned}
\dot{\rho}_{\pm}
&= R\Bigl(|\bm{E}_{2\omega}|^{4}
       + \eta_{3}^{2}\bm{E}_{2\omega}|^{2}\,|\bm{E}_{\omega}|^{4}
       + \eta_{4}^{2}\,|\bm{E}_{\omega}|^{8}\Bigr) \\
&\quad + \eta_{4}\,|\bm{E}_{2\omega}|^{2}\,|\bm{E}_{\omega}|^{4}
      \exp\bigl[2i\bigl(\phi_{2\omega}-2\phi_{\omega}\bigr)\bigr] \\
&\quad \pm \Bigl\{-\,i\,\eta_{3}\,|\bm{E}_{2\omega}|\,|\bm{E}_{\omega}|^{2}
      \bigl(\eta_{4}\,|\bm{E}_{\omega}|^{4}-|\bm{E}_{2\omega}|^{2}\bigr)
      \exp\bigl[i\bigl(\phi_{2\omega}-2\phi_{\omega}\bigr)\bigr]\Bigr\}
      + \text{c.c.},
\end{aligned}
\end{equation}
where $\bm{E}_{\omega}$ and $\bm{E}_{2\omega}$ are the local electric fields of the fundamental and second harmonic waves, respectively, $\phi_{\omega}$ and $\phi_{2\omega}$ are their phases, and $\eta_i$ represent multiphoton absorption coefficients.The first term corresponds to independent multiphoton absorption processes, involving four FW photons, two FW photons combined with one SH photon, and two SH photons. The second term describes the interaction between the four-FW and the combined two-FW–one-SH photon absorption processes. The final term represents the interaction among all three multiphoton absorption processes. It is this third term that induces asymmetric electron emission. 

And the electron trapping by hole and electron defect that create a space-charge field that grows until the photocurrent ($j_{\mathrm{ph}} \sim \dot{\rho}_{+} - \dot{\rho}_{-}$ due to anisotropic electron emssion) is balanced to the conduction current. Specially, the phase in the third term makes the DC field periodic along the wave propagation direction with period:
\begin{equation}
\Lambda = \frac{\lambda}{2(n_{2\omega}-n_{\omega})} .
\end{equation} Subsequently, an effective $\tens{\chi}^{(2)}_{\text{eff}}$ grating is formed via EFISH in centrosymmetric materials~\cite{1987Opt.Lett.StolenSelforganizedPhasematchedHarmonic,2020ACSPhotonicsNitissFormationRulesDynamics}, satisfying the quasi-phase-matching condition spontaneously~\cite{2017Opt.ExpressPorcelPhotoinducedSecondorderNonlinearity,2022Laser&PhotonicsReviewsYakarGeneralizedCoherentPhotogalvanic}.
For observable SHG via CPE, two conditions must be fulfilled: (1) initial seed SHG, potentially arising from interface states or intrinsic nonzero dipole distributions, and (2) sufficient accumulation time for the space-charge field building. 

This phenomenon was first observed in glass optical fibers~\cite{1986Opt.Lett.OsterbergDyeLaserPumped} and subsequently demonstrated in silicon~\cite{1996Phys.Rev.Lett.AtanasovCoherentControlPhotocurrenta}, GaAs~\cite{1997Phys.Rev.Lett.HacheObservationCoherentlyControlled} and perovskite~\cite{2023ACSPhotonicsZhangPhotoactivatedSecondHarmonic}systems. Direct visualization of the induced gratings was accomplished by chemically etching germanosilicate fibers using hydrofluoric acid, as the etching rate depends sensitively on the intensity of the internal electric field. Recently, silicon nitride (Si\textsubscript{3}N\textsubscript{4})~\cite{2017NatCommunBillatLargeSecondHarmonic}, characterized by ultralow optical losses and compatibility with integrated photonics and quantum technologies, has emerged as a promising platform for CPE studies. In Si\textsubscript{3}N\textsubscript{4}, nitrogen vacancies saturated by hydrogen atoms form Si--H and covalent Si--Si bonds, introducing defect states approximately $1.4$~eV away from both the conduction and valence band edges within its band gap of approximately $4.6$~eV. A multiphoton absorption mechanism involving two FW photons and one SH photon (collectively corresponding to $3.2$~eV at a wavelength of $1500$~nm) supports the hypothesis that these Si--Si defect states facilitate asymmetric carrier ejection.

The first experimental demonstration and theoretical analysis of CPE in Si\textsubscript{3}N\textsubscript{4} waveguides were presented by Adrien et~al.~\cite{2017NatCommunBillatLargeSecondHarmonic} (\textbf{Figure~8a}). They reported optically induced QPM using a continuous wave laser pump, achieving an effective nonlinear susceptibility $\chi^{(2)}$ of $0.3\,\text{pm/V}$ and a maximum SHG conversion efficiency of $0.05\%\;\text{W}^{-1}$, consistent with coherent photogalvanic theory. Variations in waveguide dimensions, however, produced differing SHG responses under identical pump conditions—a phenomenon requiring further exploration. Edgars et~al.~\cite{2020ACSPhotonicsNitissFormationRulesDynamics} addressed this issue by examining guided mode dispersion and experimentally determining the integral mode overlap factor$\kappa$, which significantly affects the properties of the $\chi^{(2)}$ grating formed by all-optical poling. Due to the coherent contributions of multiple SH modes, the resulting grating reflects a weighted superposition, with $\kappa$ determining their cumulative impact on the effective refractive index $n_{\text{eff}}^{2\omega}$ so that change the period of grating $\Lambda = \frac{\lambda}{2 |n_{\text{eff}}^{2\omega} - n_{\text{eff}}^{\omega}|}$ (\textbf{Figure~8c}).

When employing femtosecond laser pulses, spatial overlap achieved by phase matching alone is insufficient; group velocity matching between fundamental and SH pulses is crucial to sustain optimal temporal overlap. Hickstein et~al. demonstrated that the matching of group velocities minimizes temporal walkoff, allowing continuous and efficient interference between FW and SH wave, thus enhancing spatial charge separation (\textbf{Figure~8b}). This sustained temporal overlap is essential for forming stable self-organized nonlinear gratings, significantly improving SHG conversion efficiency and spectral bandwidth. Their experiments yielded a conversion efficiency of $0.005\%\;\text{W}^{-1}$ with an effective $\chi^{(2)}$ of $0.5\,\text{pm/V}$ at $1560\,\text{nm}$. Similarly, Edgars et~al.~\cite{2020Photon.Res.NitissBroadbandQuasiphasematchingDispersionengineered} demonstrated that maximizing the SHG conversion efficiency bandwidth requires minimizing wavelength dependence of the propagation constant mismatch ($\Delta\beta$). Mathematically, the condition $\partial \Delta\beta/\partial\lambda \approx 0$, equivalent to group velocity matching, achieves this optimization.

Ozan et~al.~\cite{2022Laser&PhotonicsReviewsYakarGeneralizedCoherentPhotogalvanic} further extended theoretical frameworks to include the initial phase difference ($\psi$) between seeded SH waves and the generated grating. Using slowly varying envelope and undepleted pump approximations, Maxwell's equations yield the following expression governing SH generation dynamics:
\begin{equation}
\frac{\partial^2 \bm{\bar{E}}}{\partial t\,\partial z}
\;-\;
\frac{3\,\omega\,\tens{\chi}^{(3)}}{2\,n_{2\omega}\,c\,\varepsilon}
\,\beta\,
\bigl|\bm{E}_{\omega}\bigr|^4\,
\exp\!\Bigl[i\bigl(\tfrac{\pi}{2}-\psi\bigr)\Bigr]\,
g^4(z)\,\bm{\bar{E}}
=0\,.
\end{equation}
where $\bm{\bar{E}} = \bm{E}_{2\omega} e^{t/\tau} e^{\alpha z/2}$, $\tau = \varepsilon/\sigma$ is the decay time ($\varepsilon$ is the dielectric constant, $\sigma$ is the photoconductivity), $\alpha$ denotes the SH propagation loss coefficient, and $g(z)$ represents the walk-off between the pump and SH. Initially, setting $\psi = \pi/2$ and the dominant inscription of the grating by the seeded SH wave $SH_\text{S}$, satisfying $\beta \gg \sigma$. However, the seeding SH and generated SH at $z_0$ position would be the seeding SH in $z_0+\delta z$. As the generated SH wave $SH_G$ grows and becomes comparable to the seeded SH, destructive interference between seeded and generated gratings occurs, diminishing grating formation along the waveguide. Additionally, increased photoconductivity $\sigma$ further suppresses grating longevity, limiting its effective length.This comprehensive model not only explains the phenonmon why the grating length is always limited , but also predicts the non-degenerate SFG QPM gratings (\textbf{Figure~8d}). 

To overcome limited grating lengths and enhance the effective interaction length and conversion efficiency, combining the CPE with resonant optical modes having high $Q$ factors and low mode volumes appears promising. Specifically, whispering gallery mode (WGM) microring, characterized by ultrahigh $Q$ factors and minimal mode volumes, offer excellent choice ~\cite{1986ScienceQianLasingDropletsHighlighting,1989PhysicsLettersABraginskyQualityfactorNonlinearProperties,1994Phys.Rev.Lett.LinCwNonlinearOptics,2007NaturePhysCarmonVisibleContinuousEmission,2010Phys.Rev.Lett.FurstNaturallyPhaseMatchedSecondharmonica,2014NatCommunKuoSecondharmonicGenerationUsing,2016Laser&PhotonicsReviewsBreunigThreewaveMixingWhispering,2016LightSciApplXueSecondharmonicassistedFourwaveMixing,2019NaturePhotonZhangSymmetrybreakinginducedNonlinearOptics}. The structural analogy between microring and bent waveguides shifts the mode-matching requirement to angular momentum matching. Xiyuan Wu et~al.~\cite{2021Nat.PhotonicsLuEfficientPhotoinducedSecondharmonic} first demonstrated in a Si\textsubscript{3}N\textsubscript{4} microring resonator that an initially seeded SH field and the FW generate a DC electric field via the photoelectric effect. Achieving perfect phase matching between SH and FW eigenmodes resulted in a uniform DC field (\textbf{Figure~9a}) up to $1\times10^7\,\text{V/m}$, realizing an SHG conversion efficiency of $2500\%~W^{-1}$ (absolute efficiency of 22\%) with an input power of $8.8 \pm 1.0\,\text{mW}$. Edgars et~al.~\cite{2022Nat.Photon.NitissOpticallyReconfigurableQuasiphasematching} further showed that circular polarization coherence can induce DC fields to automatically compensate for the angular momentum mismatch without precise wavelength tuning (\textbf{Figure~9b}), realizing $12.5\,\text{mW}$ on-chip SHG and a conversion efficiency of $47.6\%\;\text{W}^{-1}$. Furthermore, Jianqi et~al.~\cite{2022Sci.Adv.HuPhotoinducedCascadedHarmonic} observed analogous CPE-induced sum-frequency generation (SFG) processes in SiN microcavities, effectively yielding third-harmonic generation via FW–SH sum-frequency interactions. Both SHG and SFG gratings were experimentally demonstrated~(\textbf{Figure~9c}). Rencenty, it is found that the CPE entails a temporal oscillation to reach a stable dynamical equilibrium, which makes the SH frequency detuning with a kHz scale shift~(\textbf{Figure 8d}). 

\subsection{$\bm{E}_{\omega}$ Engineering}

The fundamental electric field distribution, $\bm{E}_\omega$, is determined by the optical modes, which are in turn governed by the symmetry of the photonic structure.  Analogous to electronic states in quantum wells, which are confined by potential barriers, optical modes in photonic resonators exhibit characteristic mode profiles dictated by their symmetry. For example, Mie theory provides an exact solution for the scattering of plane waves by a sphere, with its resonant modes described analytically by spherical harmonics (\emph{multipoles})~\cite{2019Adv.OpticalMater.BaryshnikovaOpticalAnapolesConcepts,2020Phys.Rev.BGladyshevSymmetryAnalysisMultipole}. In contrast, cylindrical resonators break spherical symmetry but retain rotational symmetry about the cylinder axis; this permits coupling among modes with the same azimuthal number and gives rise to phenomena such as quasi-BICs~\cite{2017Phys.Rev.Lett.RybinHighSupercavityModes,2019Adv.Photon.BogdanovBoundStatesContinuum,2020ScienceKoshelevSubwavelengthDielectricResonators} and exceptional points (EPs)~\cite{2025Sci.Adv.ZhangNonHermitianSingularitiesScattering}.  Moreover, two-dimensional metasurfaces—arrays of subwavelength “meta-atoms” arranged with specific lattice symmetries—have emerged as a versatile platform for controlling photonic modes. Here, the interplay between lattice symmetry and the in-plane symmetry of meta-atom determines the mode. For instance, by judiciously breaking lattice symmetry and keep the rotational symmetry using cylindrical unit cell, chiral modes can be engineered~\cite{2024Phys.Rev.Lett.ToftulChiralDichroismResonant}. And $\Gamma$ BICs can be protected by in-plane $C_2$ symmetry of the meta-atom, yielding theoretically infinite quality factors (perfect confinement without radiation loss)~\cite{2013NatureHsuObservationTrappedLight,2014Phys.Rev.Lett.ZhenTopologicalNatureOptical,2016NatRevMaterHsuBoundStatesContinuum,2018Phys.Rev.Lett.KoshelevAsymmetricMetasurfacesHigh}.  Here, we review recent advances in EFISH metasurfaces, in which resonant photonic modes are leveraged to enhance EFISH.

Despite the nonlienar metasurface inherently thin interaction region, the resonant modes derived from diverse meta-atom symmetries have enabled significant exploration and control of nonlinear phenomena. Specifically, resonance-enhanced fundamental electric fields $\bm{E}_\omega$ significantly boost SHG efficiency~\cite{2023Photon.Res.VabishchevichNonlinearPhotonicsMetasurfaces,2021PhotoniXZhaoNonlinearMetaopticsApplicationsa,2021ACSPhotonicsGrinblatNonlinearDielectricNanoantennas,2019Adv.Photon.SainNonlinearOpticsAlldielectric,2018MaterialsTodayKrasnokNonlinearMetasurfacesParadigm,2017NatRevMaterLiNonlinearPhotonicMetasurfaces,2014Rev.Mod.Phys.LapineColloquiumNonlinearMetamaterials,2023PhysicsReportsHuangResonantLeakyModes,2014NaturePhotonMeinzerPlasmonicMetaatomsMetasurfaces,2018AdvancedOpticalMaterialsRahimiNonlinearPlasmonicMetasurfaces,2021AdvancedOpticalMaterialsWangHighQPlasmonicResonances,2019ACSPhotonicsKoshelevNonlinearMetasurfacesGoverned}.

Cai et~al.~\cite{2011ScienceCaiElectricallyControlledNonlinear} first demonstrated EFISH enhancement using metallic plasmonic metagratings filled with nonlinear polymethyl methacrylate (PMMA), as illustrated in \textbf{Figure~10a}. The metallic structures defined the optical cavity and served as electrodes connected to external circuits, enabling electrical control over fields within the nonlinear medium. They achieved a modulation depth of $7\%\;\text{V}^{-1}$ and a conversion efficiency of $5.7\times10^{-11}$. A similar two-dimensional metallic metasurface (\textbf{Figure~10b}), consisting of a 50 nm thick metal layer patterned with holes on a silver substrate separated by a 100 nm alumina dielectric layer, exhibited EFISH arising from the dielectric alumina and the metal–alumina interface. This configuration yielded a modulation depth of $9\%\;\text{V}^{-1}$ and a conversion efficiency of $2\times10^{-11}$~\cite{2014NatCommunKangElectrifyingPhotonicMetamaterials}. Furthermore, employing a hybrid plasmonic electrolyte system immersed in aqueous electrolyte solution significantly enhanced EFISH due to voltage-assisted charge accumulation, achieving an exceptional modulation depth of approximately $150\%\;\text{V}^{-1}$~\cite{2016NanoLett.LanElectricallyTunableHarmonic}.

Beyond metallic structures, dielectric nanostructures offer strong optical confinement through Mie resonances~\cite{2020NanophotonicsLiuMultipoleMultimodeEngineering}. Low losses and high damage thresholds make dielectric metasurfaces attractive for nonlinear optics, especially at high pump intensities. Dielectric metasurfaces composed of high-index unit cells support high-$Q$ quasi-normal modes (QNMs) coupling to free space via multipole channels (electric dipole (ED), magnetic dipole (MD), electric quadrupole (EQ), magnetic quadrupole (MQ), etc.)~\cite{2019Adv.Photon.BogdanovBoundStatesContinuum}, enhancing harmonic generation efficiency through intensified light–matter interactions~\cite{2015Laser&PhotonicsReviewsMinovichFunctionalNonlinearOptical}. Kyu-Tae et~al.~\cite{2019ACSPhotonicsLeeElectricallyBiasedSilicon} demonstrated EFISH in a silicon grating leveraging magnetic Mie resonances, with electrodes positioned on opposite sides (\textbf{Figure~10c}). The fundamental electric field was confined within the silicon via an MD resonance, achieving a field enhancement factor of 2.4. Consequently, the SHG intensity experienced only a 1.2-fold enhancement under a 10~V external bias, corresponding to a modulation depth of approximately $1.2\%\;\text{V}^{-1}$. This modest enhancement is attributed to the MSM electrical structure design, which results in a relatively weak electrostatic field in the center of the grating.

\section{CISH and QCSE}\label{sec:CISH_QCSE}

The fundamentals of the distinction of EFISH with CISH and QCSE have been discussed in \textbf{ Section~2} in detail. Here we overview of the state of-the-art of CISH and QCSE research to distinguish it with EFISH.  

CISH, theoretically predicted by Khurgin~\cite{1995Appl.Phys.Lett.KhurginCurrentInducedSecond} and experimentally validated by van Driel et~al.~\cite{1996Phys.Rev.Lett.AtanasovCoherentControlPhotocurrenta,1997Phys.Rev.Lett.HacheObservationCoherentlyControlled} in GaAs via simultaneous one‐ and two‐photon interband absorption‐induced currents, has also been observed in several materials, including bilayer graphene (BLG)~\cite{2012NanoLett.WuQuantumEnhancedTunableSecondOrder}, silicon~\cite{2009JetpLett.AktsipetrovDCinducedGenerationReflected}, and GaAs~\cite{2012Phys.Rev.Lett.RuzickaSecondHarmonicGenerationInduced} under various electrical configurations, such as optically induced currents and external electrode biases. For example, in a metal–semiconductor–metal device formed by depositing Au electrodes on a GaAs wafer~\cite{2012Phys.Rev.Lett.RuzickaSecondHarmonicGenerationInduced} (\textbf{Figure 11a}), the measured SHG signal is proportional to the current density, in agreement with Khurgin’s theoretical description~\cite{1995Appl.Phys.Lett.KhurginCurrentInducedSecond}. In AB‐stacked graphene with an in‐plane bias applied via a dual‐gate FET~\cite{2012NanoLett.WuQuantumEnhancedTunableSecondOrder}, certain optical transitions are suppressed by a Fermi‐level shift, thereby breaking the electron distribution symmetry in momentum‐space (\textbf{Figure 11b}). In heavily doped silicon, however, earlier work~\cite{2009JetpLett.AktsipetrovDCinducedGenerationReflected} reported azimuthal anisotropy in SHG intensity similar to recent observations in graphene/SiO\textsubscript{2}/Si(001) structures, where the dominant SHG contribution arises from EFISH induced by vertical electrostatic fields at silicon interfaces rather than purely momentum‐space current‐induced effects~\cite{2013NanoLett.AnEnhancedOpticalSecondHarmonic}. This has prompted ongoing debate regarding the mechanisms responsible for CISH in different silicon devices.

Additionally, a unique type of current known as spin current represents electron distributions characterized by opposing spin momentum (\textbf{Figure~11d}). Circularly polarized photons selectively excite spin-dependent electron polarization. The resulting electron spin distributions and the selection rules governing dipole polarization elucidate the underlying chiral mechanism responsible for spin-current-induced second-harmonic generation (spin-CISH) (\textbf{Figure~11c})~\cite{2010Phys.Rev.Lett.WangSecondOrderNonlinearOptical}. Experimentally, this pure spin current can be all-optically injected into GaAs through quantum interference between one- and two-photon absorption pathways, causing electrons with opposite spins to populate states with opposite momenta~\cite{2007NaturePhysCostaAllopticalInjectionBallistica,1997Phys.Rev.Lett.HacheObservationCoherentlyControlled}. The generation of this pure spin current breaks the bulk centrosymmetry, enabling a probe field $\mathbf{E}_p$ to produce a chiral SHG signal $\mathbf{E}_\text{J}$ (\textbf{Figure~11c})~\cite{2010NaturePhysWerakeObservationSecondharmonicGeneration}.

 QCSE is typically observed in atomically thin layered materials. Due to the direct electronic band gap in the monolayer limit and strong valley-selective optical transitions in TMD, effective modulation of SHG via QCSE has been demonstrated in monolayer and few-layer MoS\textsubscript{2}~\cite{2016NanoLett.KleinStarkEffectSpectroscopy}. Notably, QCSE occurs uniformly across few-layer structures owing to negligible interlayer coupling. In contrast, EFISH-like phenomena in bilayer MoS\textsubscript{2} show substantial dependence on interlayer hybridization, significantly modifying the nonlinear optical response of individual layers~\cite{2017NanoLett.KleinElectricFieldSwitchableSecondHarmonic}.

Another atomically scaled structure exhibiting strong nonlinearities is the MQW\cite{1983IEEEJ.QuantumElectron.GurnickSyntheticNonlinearSemiconductors}. A three to five orders of magnitude greater second-order nonlinear susceptibilities ($\tens{\chi}^{(2)}$) than natural nonlinear crystals can be achieved\cite{2016AdvancedOpticalMaterialsLeeUltrathinSecondHarmonicMetasurfaces}, accompanied by significant modulation depths via the QCSE~\cite{1994IEEEJ.QuantumElectron.CapassoCoupledQuantumWell}. However, the nonlinear susceptibility tensor associated with intersubband transitions in MQWs is oriented exclusively along the growth direction, resulting in negligible nonlinear polarization for normally incident light~\cite{2025NanophotonicsKrakofskyFlatNonlinearOptics}. Consequently, subsequent research has primarily focused on harnessing intersubband nonlinearities within waveguide geometries~\cite{2002IEEEJ.QuantumElectron.CapassoQuantumCascadeLasers,2003IEEEJ.QuantumElectron.GmachlOptimizedSecondharmonicGeneration}. Recently, integrating IST–driven $\tens{\chi}^{(2)}_{\text{eff}}$ resonances in multiple quantum wells with photonic modes in metasurfaces has emerged as a powerful approach for generating harmonic radiation from free‐space excitation~\cite{2014NatureLeeGiantNonlinearResponse}. By engineering the thicknesses of wells, MQW enable precise control over subband energies and transition dipole moments~\cite{1996ScienceRosencherQuantumEngineeringOptical}, resulting in highly resonant nonlinear susceptibilities~\cite{1996ScienceRosencherQuantumEngineeringOptical,2003IEEEJ.QuantumElectron.GmachlOptimizedSecondharmonicGeneration,2007NaturePhotonBelkinTerahertzQuantumcascadelaserSource,2013NatCommunVijayraghavanBroadlyTunableTerahertz}. Lee et~al. demonstrated a giant nonlinear response (54 nm/V) by combining quantum-electronic engineering of intersubband nonlinearities with electromagnetic engineering of  L-shaped  plasmonic nanoresonators (\textbf{Figure 12a-c})~\cite{2014NatureLeeGiantNonlinearResponse}. Subsequent MQW metasurface architectures include all-dielectric bound-state-in-the-continuum designs~\cite{2022NanoLett.SarmaAllDielectricPolaritonicMetasurface}, epsilon-near-zero structures~\cite{2023AdvancedOpticalMaterialsBarbetEpsilonNearZeroEnhancementNonlinear}, and intensity-saturation–mitigating geometries~\cite{2023AdvancedMaterialsNefedkinOvercomingIntensitySaturation,2023NanoLett.KimEfficientSecondHarmonicGeneration}. Furthermore, the large Stark shifts inherent to ISTs offer additional electrical tunability of $\tens{\chi}^{(2)}$~\cite{1994IEEEJ.QuantumElectron.CapassoCoupledQuantumWell,2024AdvancedOpticalMaterialsChenElectricallyTunableStrong}, establishing MQW metasurfaces as a versatile platform for voltage-controlled nonlinear photonics. Yu et~al. designed a metasurface comprising MQW meta-atoms integrated with Au-plasmonic nanoantenna electrodes serving as bias contacts (\textbf{Figure 12d}). Under applied biases of 0, +4, and -4V, the conduction-band profile of a single MQW period changes (\textbf{Figure 12e}).Consequently, both the amplitude and phase of the effective second-order susceptibility $\tens{\chi}^{(2)}_{\text{eff}}$ are electrically modulated  (\textbf{Figure 12f}). By carefully engineering the spatial distribution of the applied voltage, a corresponding spatial profile of $\tens{\chi}^{(2)}_{\text{eff}}$ is realized, enabling dynamic control over nonlinear beam manipulations.

\section{Applications}\label{sec:applications}
\subsection{Optical Devices}

In optics, EFISH enables second-order nonlinear optical responses in centrosymmetric materials through the application of a static electric field. This approach not only circumvents the symmetry‐forbidden SHG in centrosymmetric media—thereby broadening the design space for silicon‐based integrated nonlinear photonic devices—but also enables electrical control over the spatial modulation of the effective second‐order susceptibility \(\tens{\chi}^{(2)}_{\mathrm{eff}}\), making EFISH a highly promising effect for efficient and electrically tunable SHG.The performance metrics for second-harmonic generation across various material and structural configurations are comprehensively summarized in \textbf{Table~1}.

Specifically, by spatially modulating the effective second-order nonlinear susceptibility \(\tens{\chi}^{(2)}_{\mathrm{eff}}\), EFISH permits optical or electrical tuning of phase-matching conditions between the fundamental and second-harmonic waves, thereby opening new opportunities for integrated nonlinear photonic devices such as high fruquency laser~\cite{2021Nat.PhotonicsJinHertzlinewidthSemiconductorLasers,2023Nat.Photon.Corato-ZanarellaWidelyTunableNarrowlinewidth,2021ScienceXiangLaserSolitonMicrocombs}, optical parametric amplification (OPA) waveguide and frequency-comb self-referencing~\cite{2018NatureSpencerOpticalfrequencySynthesizerUsing,2024NatCommunZhangHighcoherenceParallelizationIntegrated,2025NatElectronZhangMicrocombsynchronizedOptoelectronics}. For example, OPA requires precise phase engineering of the pump, signal and idler waves~\cite{baumgartner1979optical}. EFISH-induced periodic $\tens{\chi}^{(2)}$ gratings can engineer the momentum matching~\cite{2017NaturePhotonTimurdoganElectricFieldinducedSecondorder}, as detailed in Section~2.2. Thus provide a promising platfom for high performance OPA. Heydari \emph{et al.}~\cite{2023OpticaHeydariDegenerateOpticalParametric} employed this periodic modulation of the effective second-order susceptibility \(\bm{\chi}^{(2)}_{\mathrm{eff}}\) in silicon waveguides by integrating alternating p--i--n diode arrays (\textbf{Figure 13a}). This spatially varying bias profile enables quasi-phase-matched OPA directly on-chip. Under strong second-harmonic pumping at 1196\,nm with a 31\,V reverse bias, their device achieved phase-sensitive amplification at the fundamental wavelength, reaching a normalized gain of $0.6\,\mathrm{dB}\,\mathrm{W}^{-1/2}\,\mathrm{cm}^{-1}$ .

While electrodes design provide spatially uniform DC fields with high freedom, complex fabrication is required. CPE offers an all-optical alternative by inscribing periodic DC-field–induced \(\tens{\chi}^{(2)}_{\mathrm{eff}}\) gratings directly into photonic structures—such as straight waveguides and microring resonators satisfying the quasi-phase-matching condition spontaneously. employing a dual-resonator CPE architecture, conversion efficiencies up to \(2500\,\mathrm{W}^{-1}\) with milliwatt-level SHG output have been demonstrated~\cite{2021Nat.PhotonicsLuEfficientPhotoinducedSecondharmonic}. The CPE-inscribed grating (approximately 4 µm in period) enables the QPM to support \(\sim100\,nm\) of bandwidth; however, perfect phase matching is restricted to a single resonant mode, thereby constraining the usable spectral range. To address this, Marco \emph{et al.}~\cite{2025NatCommunClementiUltrabroadbandMilliwattlevelResonant} using separate racetrack resonators for the FH and SH (the South and North rings in \textbf{Figure~12c}), respectively. A Mach–Zehnder interferometer (MZI) with tunable thermal phase shifters decouples FW from north ring (achieving a \(100:0\) splitting ratio in the pump band) and SH entirely into the south ring (\(0:100\)). Both resonators were engineered to satisfy free spectral range (FSR) matching \(\mathrm{FSR}_{\rm FW}\approx\mathrm{FSR}_{\rm SH}\), enabling multiple doubly resonant mode pairs across the QPM bandwidth. Within their shared waveguide segment, CPE inscribes a periodic DC‐field–induced \(\chi^{(2)}_{\mathrm{eff}}\) grating that meets the QPM condition over more than 200\,nm of spectral range. As a result, this dual‐resonator system generates broadband, milliwatt‐level SHG in the North ring with \(40\%~\mathrm{W}^{-1}\) conversion efficiency. 

Additionally, Marco \emph{et al.}~\cite{2023LightSciApplClementiChipscaleSecondharmonicSource} integrated an electrically pumped distributed-feedback (DFB) laser with a \(\mathrm{Si}_3\mathrm{N}_4\) microring resonator realizing a self-injection-locked (SIL) second-harmonic source with an ultranarrow emission linewidth \textbf{(Figure~12b)}. In their design, the DFB laser is edge-coupled into the microring, where CPE inscribes a DC-field-induced \(\tens{\chi}^{(2)}\) grating in the fundamental transverse-electric (TE\(_{00}\)) mode to achieve quasi-phase matching. Simultaneously, Rayleigh backscattering from sidewall roughness provides intrinsic optical feedback by recirculating a fraction of the microring field back into the laser cavity, suppressing phase noise by a factor proportional to \(Q^2\). Because the SIL-locked emission frequency coincides with the TE\(_{00}\) resonance, the same inscribed \(\tens{\chi}^{(2)}\) grating automatically satisfies the quasi-phase-matching condition for second-harmonic generation. This dual-function architecture—combining SIL linewidth stabilization with on-chip EFISH-enabled strong sidemode suppression ratio (SMSR) exceeding \(60~dB\).

Recently, this preeminent second‐order nonlinearity property in Si$_3$N$_4$ microring resonators has been developed for spontaneous parametric down‐conversion (SPDC) (\textbf{Figure~12d})~\cite{2025NatureLiDownconvertedPhotonPairsa}. In the reported scheme, the fundamental transverse‐electric mode at 1560\,nm (chosen for optimal CPE) is pumped to inscribe a long‐lived DC‐field–induced $\tens{\chi}^{(2)}_{eff}$ grating via CPE. Upon subsequent injection of a second‐harmonic pump, this preexisting grating satisfies the quasi‐phase‐matching condition for SPDC. Resonant enhancement at down‐conversion wavelength by the microring, together with the large FSR of the resonator, leads to entangled photon‐pair generation rates up to 0.8\,million pairs per second.

\subsection{Probing of Carriers Dynamics}

The different mobility of electron and hole generates a static electric field that enables EFISH, as discussed in Section~2.2. Conversely, EFISH signals can serve as effective optical probes for carrier dynamics, a capability essential for both electronic~\cite{2007NaturePhotonManakaDirectImagingCarrier,2015LightSciApplManakaOpticalSecondharmonicGeneration}, optoelectronic~\cite{2009Phys.Rev.Lett.DevizisUltrafastDynamicsCarrier} and THz~\cite{2020AdvancedOpticalMaterialsLeeElectricallyControllableTerahertz} devices.

Most electronic devices operate via the spatial transport of charge carriers. For example, FETs exploit carrier motion in a semiconductor channel that is modulated by an external gate voltage~\cite{2005Appl.Phys.Lett.ManakaModulationOpticalSecond,2006jjapLimMaxwellWagnerModel,2006J.Appl.Phys.TamuraAnalysisPentaceneFielda}. In inorganic semiconductors, such as silicon~\cite{2004NatureXiaoElectricalDetectionSpin,2017NatureAdinolfiPhotovoltageFieldeffectTransistors}, well-established experimental methods and physical models have promoted rapid advancements. Conversely, the underlying physics of organic~\cite{1998Adv.Mater.HorowitzOrganicFieldEffectTransistors} and ferroelectric~\cite{2020NatElectronKhanFutureFerroelectricFieldeffect} electronic devices remains less well understood, hindering performance optimization. Conventional scanning probe methods, such as scanning Kelvin probe microscopy and AFM-based potentiometry, provide high-resolution scalar potential mapping but lack direct measurement capabilities of vectorial electric fields. EFISH provides direct, non-invasive optical imaging of the in-plane vector electric field with approximately $0.5\,\mu\text{m}$ spatial and $5\,\text{ns}$ temporal resolutions~\cite{2007NaturePhotonManakaDirectImagingCarrier}. The coherent nature of SHG—highly sensitive to both the magnitude and orientation of static fields via the nonlinear susceptibility tensor—makes EFISH ideal for visualizing carrier injection, transport, and accumulation. Consequently, EFISH complements traditional scanning probe and charge modulation techniques by offering enhanced spatiotemporal resolution, critical for optimizing organic device performance~\cite{2009Appl.Phys.Lett.TaguchiProbingCarrierBehavior,2009J.Appl.Phys.WeisOriginElectricField,2010J.Phys.Chem.Lett.TaguchiAnalysisOrganicLightEmitting}. Furthermore, time-resolved EFISH offers direct insights into carrier dynamics, which are essential for improving device switching speeds, reliability, and efficiency. For example, Xu et~al.~\cite{2024NatCommunXuEnhancementsElectricField} utilized time-resolved EFISH to study transient electric fields in ferroelectric barrier discharge (FBD) systems employing Pb(Zr$_x$Ti$_{1-x}$)O$_3$ (PZT) electrodes. They observed field overshoots (approximately $13\,\text{kV/cm}$ compared to $5\,\text{kV/cm}$ in conventional discharges) and significantly prolonged plasma durations due to rapid ferroelectric polarization-induced surface charges, phenomena highly relevant for plasma catalysis and material synthesis.

In optoelectronic devices, carrier dynamics at ultrafast timescales are critically important. Unlike electrically driven carriers whose response is inherently limited by comparatively slow electrical modulation speeds, optically excited carriers respond on femtosecond timescales. Leveraging ultrafast spectroscopic techniques, EFISH can directly probe these ultrafast carrier dynamics with femtosecond temporal resolution. Specifically, in centrosymmetric semiconductors, time-resolved EFISH (TREFISH) effectively measures interfacial electric fields arising from ultrafast electron transfers, exhibiting exceptional sensitivity at crystalline interfaces due to the intrinsic bulk-forbidden SHG signal. Devizis et~al.~\cite{2009Phys.Rev.Lett.DevizisUltrafastDynamicsCarrier} first demonstrated TREFISH in conjugated polymer-based devices employing ITO electrodes (\textbf{Figure~13a}), revealing charge relaxation dynamics within disordered organic solids. Subsequently, TREFISH became instrumental in studying photovoltaic mechanisms, notably within organic photovoltaics (OPVs)~\cite{2013NatCommunVithanageVisualizingChargeSeparation} and hot-electron photovoltaic~\cite{2010ScienceTisdaleHotElectronTransferSemiconductor} devices.

Typical photovoltaic processes include~\cite{nelson2003physics}: (i) photogeneration of hot electrons by photons whose energies exceed the semiconductor bandgap; (ii) rapid cooling resulting in exciton formation (electron--hole pairs bound by Coulomb attraction), necessitating subsequent separation at donor--acceptor interfaces; and (iii) charge transfer processes yielding free carriers. However, carrier dynamics in organic materials are inherently more complex, which limits the power‐conversion efficiency of OPVs compared to their inorganic counterparts~\cite{2023NatRevMaterYiAdvantagesChallengesMolecular}. In 2013, Vithanage et~al.~\cite{2013NatCommunVithanageVisualizingChargeSeparation} utilized TREFISH to elucidate charge-separation mechanisms in bulk-heterojunction organic solar cells, revealing rapid spatial separation (several nanometers within picoseconds) of initially closely bound electron--hole pairs ($<1\,\text{nm}$). This rapid separation overcome Coulombic attraction, generating free carriers within sub-nanosecond timescales (\textbf{Figure~14b}). Insights from such studies guided the development of advanced electron-donor and electron-acceptor materials, including polymers and fullerenes. Conventionally, carrier mobilities in organic materials are treated as near-equilibrium constants determined by diverse measurement techniques. However, Melianas et~al.~\cite{2014AdvFunctMaterialsMelianasDispersionDominatedPhotocurrentPolymer}, combining TREFISH experiments with kinetic Monte Carlo simulations, demonstrated significant time dependence in carrier mobility (\textbf{Figure~14c}). Moreover, the inherently low dielectric constant of organic semiconductors intensifies electron--hole Coulomb interactions, complicating free-carrier generation. While the exact mechanism allowing electrons and holes to escape Coulombic trapping remains incompletely understood, mounting evidence highlights the critical role of hot charge-transfer (CT) excitons in facilitating free-carrier formation. Jailaubekov et~al.~\cite{2013NatureMaterJailaubekovHotChargetransferExcitons}employed femtosecond EFISH measurements to directly visualize hot CT exciton formation and relaxation dynamics in OPVs (\textbf{Figure~14d}).

Beyond traditional electron-hole pair separation, directly harvesting hot electrons before thermalization offers promising pathways for efficiency improvement. Typically, hot electrons rapidly lose energy through ultrafast relaxation processes. As previously discussed, hot-electron dynamics generated via LSPR in metallic structures can be probed through ultrafast EFISH signals~\cite{2020Phys.Rev.Lett.TaghinejadTransientSecondOrderNonlinearb}. Analogously, EFISH is effective in examining hot-electron dynamics in semiconductor systems. One promising approach involves semiconductor nanocrystals or quantum dots (QDs)~\cite{alivisatos1996semiconductor}, where quantum confinement discretizes electronic energy bands. Such discretization often results in energy-level spacings greater than lattice phonon energies, creating a ‘phonon bottleneck’ that significantly slows hot-electron relaxation via multiphonon processes. Tisdale et~al.~\cite{2010ScienceTisdaleHotElectronTransferSemiconductor} employed SHG signals to observe hot-electron transfer from colloidal PbSe nanocrystals to TiO$_2$, resolving transfer events on the order of $\sim50\,\text{fs}$ (\textbf{Figure~14e}). Similarly, Williams et~al.~\cite{2013ACSNanoWilliamsHotElectronInjection} investigated hot-electron transfer dynamics from graphene quantum dots to TiO$_2$ (\textbf{Figure~14f}). Their study identified two distinct decay pathways: (1) rapid recombination of interfacial electron-hole pairs occurring within the initial $15\,\text{fs}$, and (2) a slower, temperature- and excitation-energy-dependent relaxation pathway termed the ``boomerang'' mechanism. In the latter mechanism, hot electrons injected into bulk TiO$_2$ undergo cooling via electron--phonon interactions, then drift back toward the interface under transient electric fields, recombining with holes on the graphene quantum dots within approximately $2\,\text{ps}$.

Another intriguing area of research involves the response at terahertz (THz) frequencies, situated between the microwave and optical regimes. Unlike the dipole generating by electron-cloud distortion with optical incident, or steady-state carrier drift described typically by current–voltage (I–V) characteristics, the THz response is dominated by ultrafast displacement of free or weakly bound carriers~\cite{lee2009principles}. This displacement involves time-dependent processes such as carrier acceleration, scattering, and transient redistribution. Therefore, the EM response at THz frequencies are typically characterized by the frequency-dependent permittivity $\varepsilon(\omega)$,which describe dipole generating by carriers movement. In most materials and devices studied so far, carriers drift symmetrically along the direction of the incident THz electric field, resulting in nonlinear optical responses predominantly described by odd-order nonlinear susceptibilities. Lee et~al.~\cite{2020AdvancedOpticalMaterialsLeeElectricallyControllableTerahertz} overcame symmetry limitations by engineering a GaAs structure with interdigitated electrodes. An incident optical pump excited electrons from the valence band to the conduction band, while intervalley scattering in doped GaAs provided large third-order nonlinear optical interactions at THz frequencies~\cite{2013Phys.Rev.Lett.FanNonlinearTerahertzMetamaterials}. The external bias field, applied via these comb-like electrodes, disrupted the symmetry of photocarrier diffusion. Consequently, under optical excitation, this induced asymmetry in carrier distribution produced an observable THz EFISH signal (\textbf{Figure 15a}).

In addition to driving carrier transport, intense THz radiation can also induce polarization switching in ferroelectric materials via resonant control over ionic motion as well as lattice vibrations. As previously discussed in Section~2.1, polarization-state switching induced by external bias fields strongly modulates SHG~\cite{2021AdvancedOpticalMaterialsFeutmbaReversibleTunableSecondOrder}. EFISH can similarly serve as an optical probe to investigate optically induced ultrafast polarization-switching dynamics in ferroelectrics. Chen et~al.~\cite{2015AdvancedMaterialsChenUltrafastTerahertzGating} employed EFISH to probe these ultrafast dynamics. They fabricated EO devices incorporating interdigitated electrode structures (\textbf{Figure~15b}) and recorded the SHG signal under 800\,nm optical excitation. \textbf{Figure~15c} shows the SHG response of a BiFeO$_3$ thin film within this EO structure. At low bias fields, the SHG intensity exhibits the characteristic EFISH modulation; when the applied field exceeds the known coercive field, pronounced “butterfly” loops appear, indicating full ferroelectric polarization switching, coincident with the recent result~\cite{2021AdvancedOpticalMaterialsFeutmbaReversibleTunableSecondOrder}. Moreover, by driving the film with single‑cycle THz pulses, they generated transient electric fields on a sub‑picosecond timescale, thereby modulating the polarization state. For small incident THz fields, the measured EFISH modulation has the same temporal dependence as the electric field as measured by electro-optic sampling~(\textbf{Figure~15d}), indicate the low field THz drives an adiabatic, reversible tilt of the polarization vector rather than a hysteretic, permanent switching.

\section{Summary and Outlook}

As demonstrated above, EFISH transcends the inherent symmetry limitations of traditional nonlinear material, opening new avenues for applications across electrical~\cite{2015LightSciApplManakaOpticalSecondharmonicGeneration}, optical~\cite{2025NatureLiDownconvertedPhotonPairsa}, optoelectronic~\cite{2015Phys.Chem.Chem.Phys.BasslerHotColdHow}, and terahertz~\cite{2020AdvancedOpticalMaterialsLeeElectricallyControllableTerahertz} devices. Recent pioneering work in material synthesis and structural engineering has demonstrated that the distortion of electron clouds under electrostatic fields—coupled with electron-based quasiparticles such as skyrmions~\cite{2024NatCommunWangGiantElectricFieldinduceda}, excitons~\cite{2017NanoLett.KleinElectricFieldSwitchableSecondHarmonic}, and plasmons~\cite{2024J.Appl.Phys.GoswamiExploringSynergyHotelectron}—can dramatically enhance the EFISH response. The photonics structure design including both localized and resonant leaky photonic states~\cite{2011ScienceCaiElectricallyControlledNonlinear} have also shown superior performance in EFISH, underscoring the technique’s versatility.

The integration of advanced photonic mode designs with tailored electronic properties is expected to further propel the capabilities of EFISH. In particular, nanophotonic platforms that exploit LSPR in metals~\cite{2021AdvancedOpticalMaterialsWangHighQPlasmonicResonances} and Mie resonances~\cite{2016ScienceKuznetsovOpticallyResonantDielectric,2024Nat.Photon.ZografCombiningUltrahighIndexa} in dielectrics offer exceptional light confinement and manipulation at the nanoscale. This synergy positions EFISH as a highly promising strategy for developing next-generation nonlinear optical devices and tunable photonic systems. Nevertheless, realizing this potential demands a careful co-design approach that harmonizes both photonic and electronic functionalities. A key challenge remains the design of electric elements at the nanoscale, in particular the incorporation of static electric fields into the nanoscale structure.

Moreover, the time-resolved EFISH response provides an exceptional platform for high spatial and temporal accuracy in probing carrier dynamics~\cite{2009Phys.Rev.Lett.DevizisUltrafastDynamicsCarrier}. While the optical polarization response occurs on timescales as short as attoseconds to femtoseconds, carrier dynamics—including excitation, relaxation, and recombination—span from femtoseconds to nanoseconds. Longer processes such as thermalization, phonon interactions, and structural phase transitions may even extend into the microsecond regime. This ultrafast probe capability makes EFISH an invaluable tool for characterizing complex carrier dynamics in ferroelectrics~\cite{2018AdvancedMaterialsZhangFerroelectricPiezoelectricEffects}, donor-acceptor systems~\cite{2011Chem.Mater.EllingerDonorAcceptorDonor}, 2D ferroelectrics~\cite{2021AdvancedMaterialsQiReviewRecentDevelopments,2021ACSNanoWuTwoDimensionalVanWaals,2023Nat.Mater.WangTwodimensionalVanWaals} and 2D metal~\cite{2025NatureZhaoRealization2DMetals}.

We believe that the evolving landscape of the EFISH research not only broadens our fundamental understanding of light–matter interaction but also paves the way for innovative optoelectronics applications. Continued advancements in integrated photonic and electronic design will advance this transformative technology.


\medskip
\textbf{Acknowledgments} \par 
The authors are thankful to Mikhail Lapine and Wenshan Cai for the fruitful discussions. A part of this work was supported by the
Australian Research Council (Grant No. DP210101292)

\newpage

\medskip

%

\bibliographystyle{MSP} 
\bibliography{Ref}

\newpage


\newpage

\begin{figure}
    \centering
    \includegraphics[width=\linewidth]{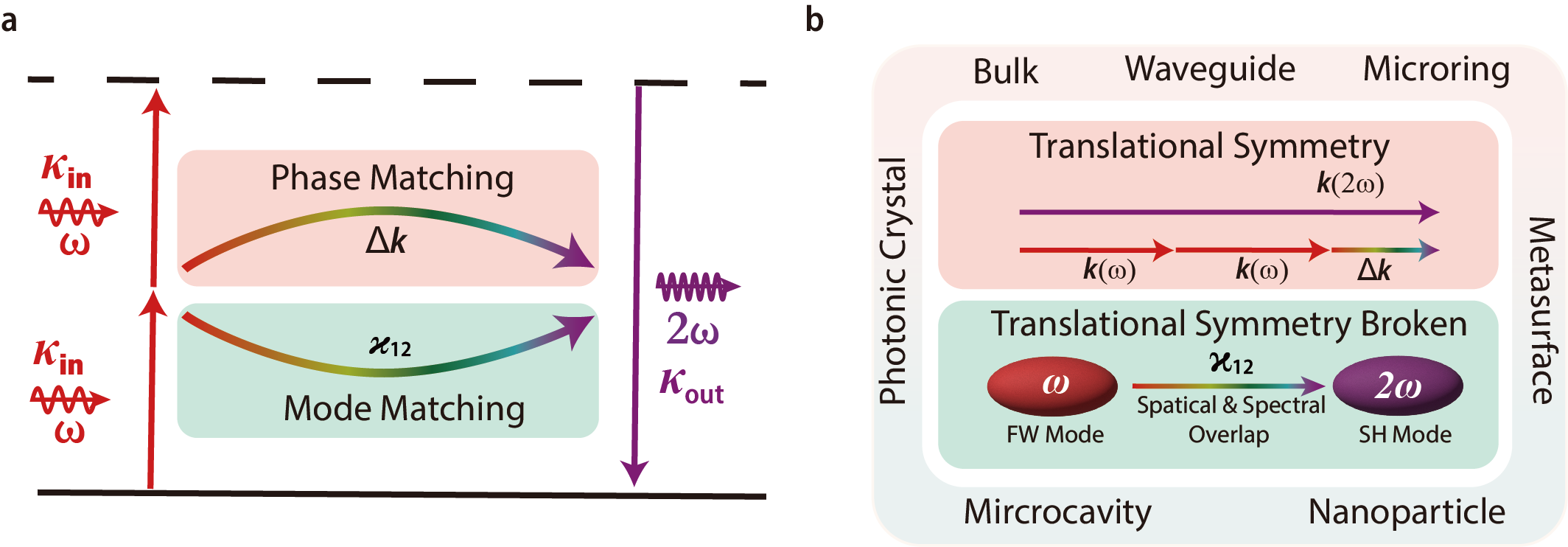}
    \caption{\textbf{Fundamentals of SHG.} 
    a)~Schematic energy-level diagram illustrating SHG. Efficient nonlinear harmonic generation requires both phase matching and mode matching. 
    b)~Illustration of phase versus mode matching based on translational symmetry. In systems with preserved translational symmetry (e.g., bulk media, waveguides and microring), momentum is a well-defined quantum number and phase matching is essential. When translational symmetry is broken (e.g., unitcell of metasurfaces~\cite{2015NatureNanotechCelebranoModeMatchingMultiresonant, 2015NatureMaterOBrienPredictingNonlinearProperties}, photonic crystals defect cavities~\cite{2019OpticaMinkovDoublyResonantH2}, and nanoparticle~\cite{2020ScienceKoshelevSubwavelengthDielectricResonators}), momentum conservation is relaxed and mode matching governs nonlinear photon interactions.
}
    \label{fig:fundamentals}
\end{figure}

\begin{figure}
    \centering
    \includegraphics[width=\linewidth]{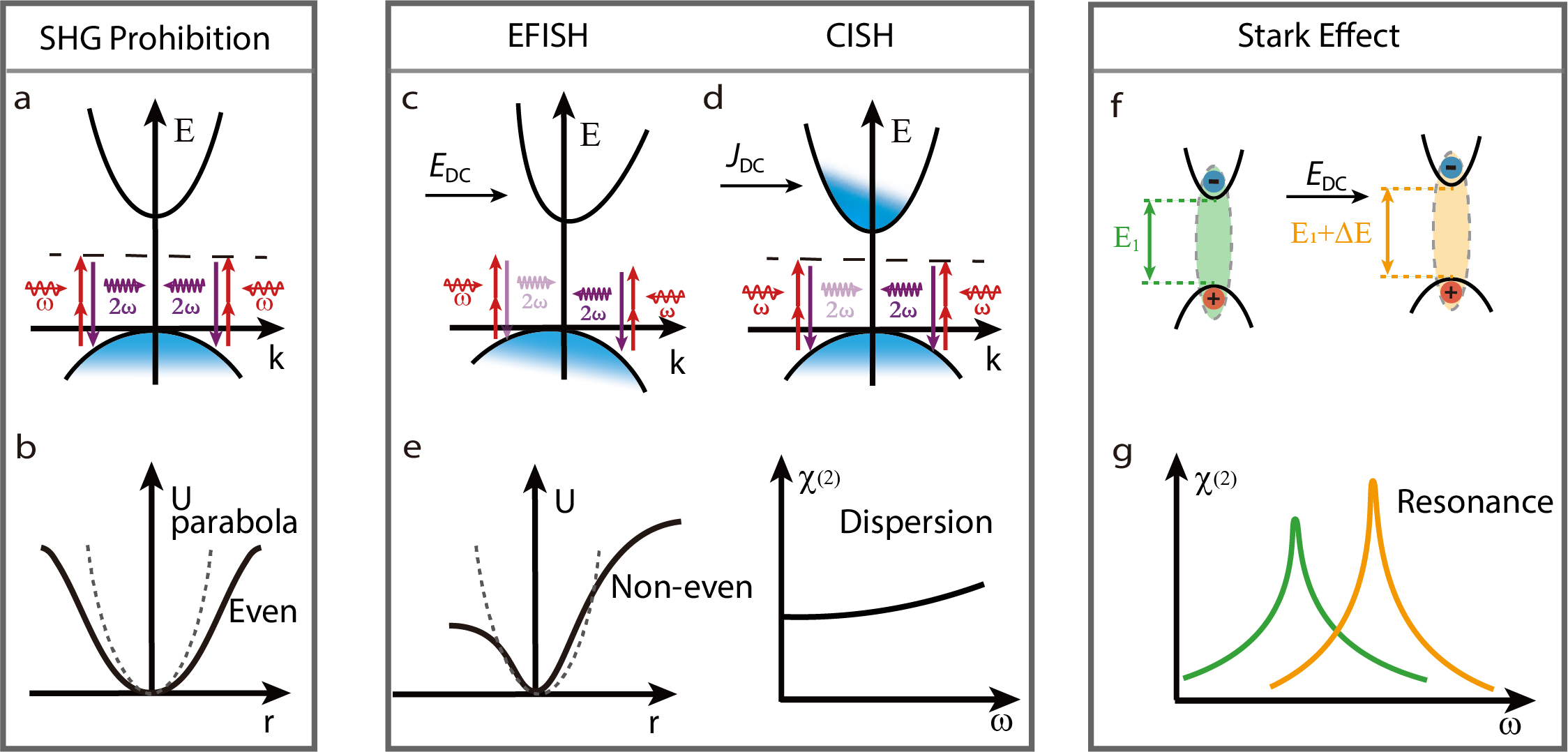}
    \caption{\textbf{Distinct mechanisms of EFISH, CISH, and QCSE.}
    a, b)~Mechanism illustrating SHG suppression in centrosymmetric materials. In such crystals, symmetric electron distributions (blue area) and potentials in momentum space (a) correspond to even potentials in real space (b), resulting in mutually cancelling second-harmonic polarizations, thereby forbidding SHG.
    c)~Application of an external electric field breaks centrosymmetry by inducing asymmetric electron potentials between momentum states at $\mathbf{k}$ and $-\mathbf{k}$ (EFISH), thus creating an effective second-order nonlinear susceptibility $\chi^{(2)}_\text{eff}$.
    d)~In the material with high carrier density, charge carriers movement (current) breaks symmetry through asymmetric electron distributions in momentum space, giving rise to CISHG with an effective $\chi^{(2)}_\text{eff}$.
    e)~Both EFISH and CISHG correspond to asymmetric potentials in real space with dispersion of the induced second-order nonlinear susceptibility.
    f, g)~Formation of excitons or intersubband transitions leads to resonant enhancement in $\chi^{(2)}$. Application of a static electric field alters the electronic band structure via the Stark effect, thereby modulating this resonant $\chi^{(2)}$.
}
    \label{fig:mechanisms}
\end{figure}

\begin{figure}
  \includegraphics[width=\linewidth]{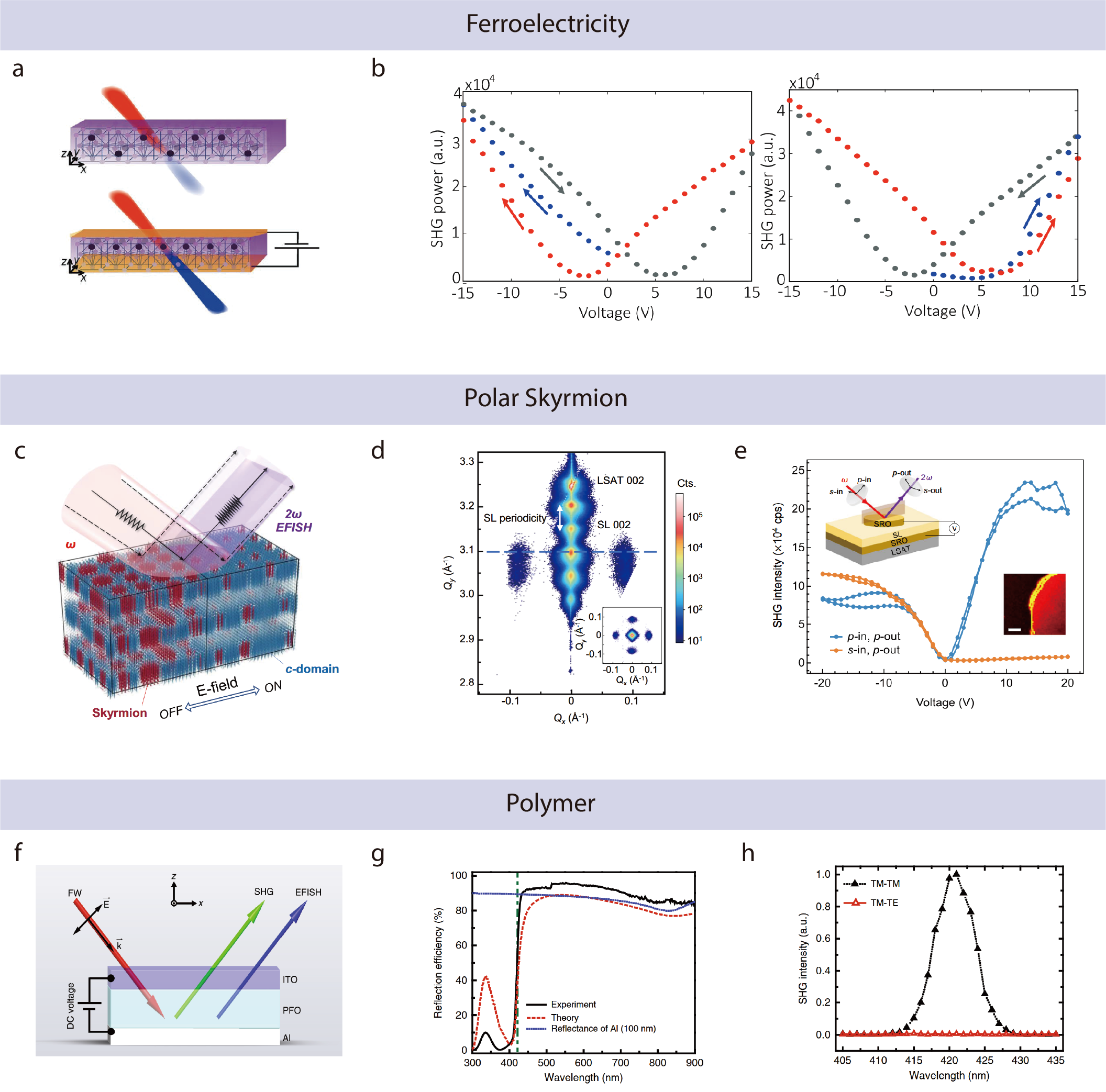}
  \caption{\textbf{Bulk nonlinear material engineering for EFISH.} 
  a) Schematic of the PZT thin film poled along the $z$‑axis and the trajectories of the incident fundamental beam ($\omega$, red) and the SH beam ($2\omega$, blue). Black dots denote $Ti/Zr$ ions, which define the film’s polarization density. 
  b) Evolution of SHG power under a triangular DC bias: starting from 0V, the bias follows the blue, gray, then red curves. The observed butterfly hysteresis in SHG confirms ferroelectric domain alignment in the PZT film.
  Reproduced with permission.~\cite{2021AdvancedOpticalMaterialsFeutmbaReversibleTunableSecondOrder} Copyright 2021, Wiley-VCH GmbH.
  c) Phase‑field model schematic of an electric‑field–induced transition in the polarization‑vector configuration of polar skyrmions. 
  d) H0L slice ($Q_{y}=0\,\mathring{A}^{-1}$) X‑ray reciprocal‑space map around the LSAT(substrate) 002 reflection for a [PbTiO$_3$]$_{14}$/[SrTiO$_3$]$_{16}$ superlattice; the arrow indicates superlattice periodicity along $Q_z$. Inset: HK0 slice ($Q_z=3.097\,$\AA$^{-1}$) showing in‑plane skyrmion periodicity with four lobes along H00/0K0. 
  e) Measured SHG intensity versus applied voltage for p‑ (parallel) and s‑ (perpendicular) polarized fundamental light. Bias sequence: $0\rightarrow 20\rightarrow-20\rightarrow0\,V$. Insets: 45\,$\deg$ measurement geometry for in situ SHG on a SrRuO$_3$/SL/SrRuO$_3$ capacitor; SHG image near the electrode boundary under normal incidence.
  Reproduced with permission.~\cite{2024NatCommunWangGiantElectricFieldinduceda} Copyright 2021, Nature Publishing Group.
  f) Schematic of EFISH from the ITO/PFO/Al device. 
  g) Measured reflection spectra of the EFISH device. The reflection efficiency sharply decreases below 430\,nm due to PFO absorption.
  h) Polarization dependence of SHG at 420\,nm: spectra for TM (same as fundamental) and TE (cross polarized along $y$) states. The TM polarized SHG signal is significantly stronger than the TE component. Reproduced with permission.~\cite{2019LightSciApplChenGiganticElectricfieldinducedSecond} Copyright 2019, Nature Publishing Group.}
  \label{fig:boat2}
\end{figure}

\begin{figure}
  \includegraphics[width=\linewidth]{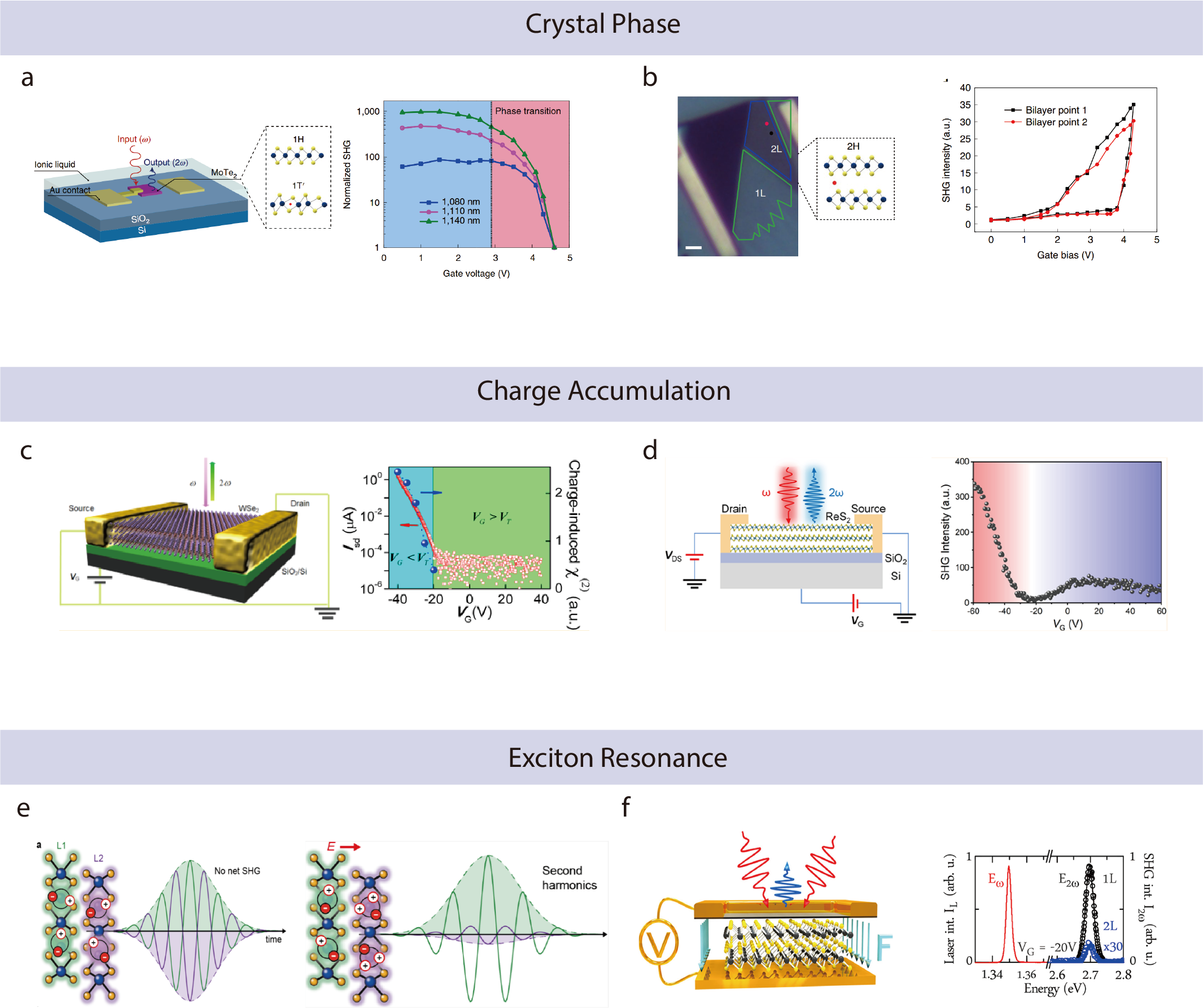}
  \caption{\textbf{Electric‑field–induced symmetry breaking in vdW materials.} Crystal Phase:
    a) Schematic of an SHG modulator: ionic liquid DEME–TFSI provides electrostatic doping under gate bias $V_{\mathrm{g}}$, driving a structural transition between the 1H and 1T$'$ phases in monolayer MoTe$_2$. The inversion center (red dot) is present in the 1T$'$ phase. 
    b) Optical micrograph of a mixed monolayer (1L, green outline) and bilayer (2L, blue outline) MoTe$_2$ flake contacted by electrodes (yellow rectangles). Gate‑dependent SHG curves from two bilayer regions show opposite trends, with a sharp increase above 3.8\,V, indicating inversion symmetry breaking. The excitation wavelength was fixed at 1060\,nm. The observed hysteresis accompanies the structural change in the bilayer.
    Reproduced with  permission.\cite{2021NatElectronWangDirectElectricalModulation} Copyright 2021, Nature Publishing Group.
    Charge Accumulation:
    c) Schematic of the bilayer WSe$_2$ device configuration and $I_{\mathrm{sd}}$–$V_{\mathrm{g}}$ characteristics. Region I (blue) corresponds to $V_{\mathrm{g}}<V_{\mathrm{T}}$ (hole accumulation) and region II (green) to $V_{\mathrm{g}}>V_{\mathrm{T}}$. SHG intensities (blue dots) are overlaid on the $I_{\mathrm{sd}}$–$V_{\mathrm{g}}$ curve. Reproduced with  permission.~\cite{2015NanoLett.YuChargeInducedSecondHarmonicGeneration} Copyright 2015, American Chemical Society.
    d) Structure diagram of a trilayer ReS$_2$ FET. Excitation at $\omega$ (red arrow) generates SHG at $2\omega$ (blue arrow). Gate‑voltage dependence of SHG intensity in the trilayer device. Reproduced with  permission.~\cite{2022ACSNanoWangElectricallyTunableSecond} Copyright 2022, American Chemical Society.
    Excition resonance:
    e) In 2H‑stacked bilayer WSe$_2$, intralayer exciton states in each layer produce opposite SHG signals that cancel, yielding zero net SHG. Selective hole injection into one layer induces exciton–polaron states, while the other layer remains charge neutral, resulting in distinct resonances and enhanced SHG. Reproduced with  permission.~\cite{2024NanoLett.ChaEnhancingResonantSecondHarmonic} Copyright 2024, American Chemical Society.
    f) Schematic of a bilayer MoS$_2$ microcapacitor with a semitransparent top gate for simultaneous optical access and gate bias $V_{\mathrm{G}}$. Typical SHG spectra at $E_{\omega}=1.35\,$eV for monolayer (black circles) and bilayer (blue circles, 30$\times$ magnified) 2H‑MoS$_2$ at $V_{\mathrm{G}}=-20\,$V, showing restored inversion symmetry in the bilayer. Reproduced with  permission.~\cite{2017NanoLett.KleinElectricFieldSwitchableSecondHarmonic} Copyright 2017, American Chemical Society.}
  \label{fig:vdW_SHG}
\end{figure}

\begin{figure}
  \includegraphics[width=\linewidth]{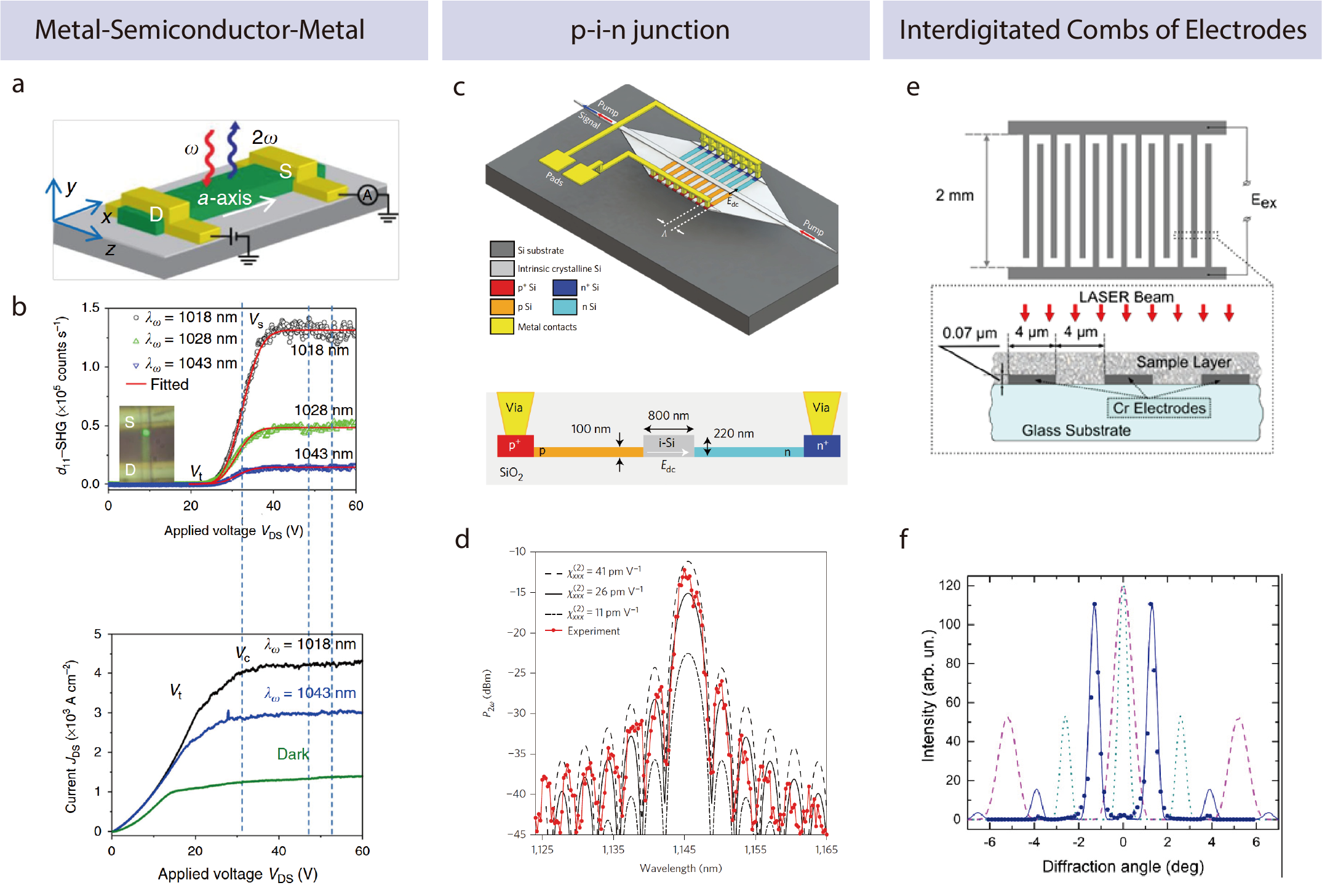}
  \caption{\textbf{Electrostatic field engineering in EFISH.} Metal-semiconductor-metal:
    a) Schematic of the CdS nanobelt device with source (S) and drain (D) electrodes.
    b) Position dependence of the $d_{11}$ SHG signal. Inset: optical micrograph showing the FW incident near the grounded source electrode (S, 0V) while the drain electrode (D) is biased. Under positive $V_{\mathrm{DS}}$, current flows along the $x$‑axis from D to S, reverse‑biasing the S contact.d11-SHG signal scanned over the entire nanobelt from the source to drain electrodes at VDS = 30 V. Reproduced with  permission.~\cite{2018NatCommunRenStrongModulationSecondharmonic} Copyright 2018, Nature Publishing Group.
    p-i-n junction:
    c)Schematic of EFISHG device with silicon ridge waveguide and quasi-phase-matched spatially periodic patterning of the p–i–n junction.
    d) On‑chip SHG intensity curves for estimated second‑order susceptibilities. Reproduced with  permission.~\cite{2017NaturePhotonTimurdoganElectricFieldinducedSecondorder} Copyright 2017, Nature Publishing Group.
    Interdigitated combs of electrodes:
    e) Schematic of interdigitated electrode combs: top‑view and side‑view. $E_{\mathrm{ex}}$ denotes the applied external electric field.
    f) Angular dependence of the EFISH intensity: experimental data (points) and calculation (solid blue line). Dashed magenta and dotted green lines show the calculated Fraunhofer diffraction patterns for the fundamental and second‑harmonic signals, respectively. Reproduced with  permission.~\cite{2016Opt.Lett.JasinskasBackgroundfreeElectricFieldinduced} Copyright 2016, Optica Publishing Group. }
  \label{fig:boat4}
\end{figure}

\begin{figure}
  \includegraphics[width=\linewidth]{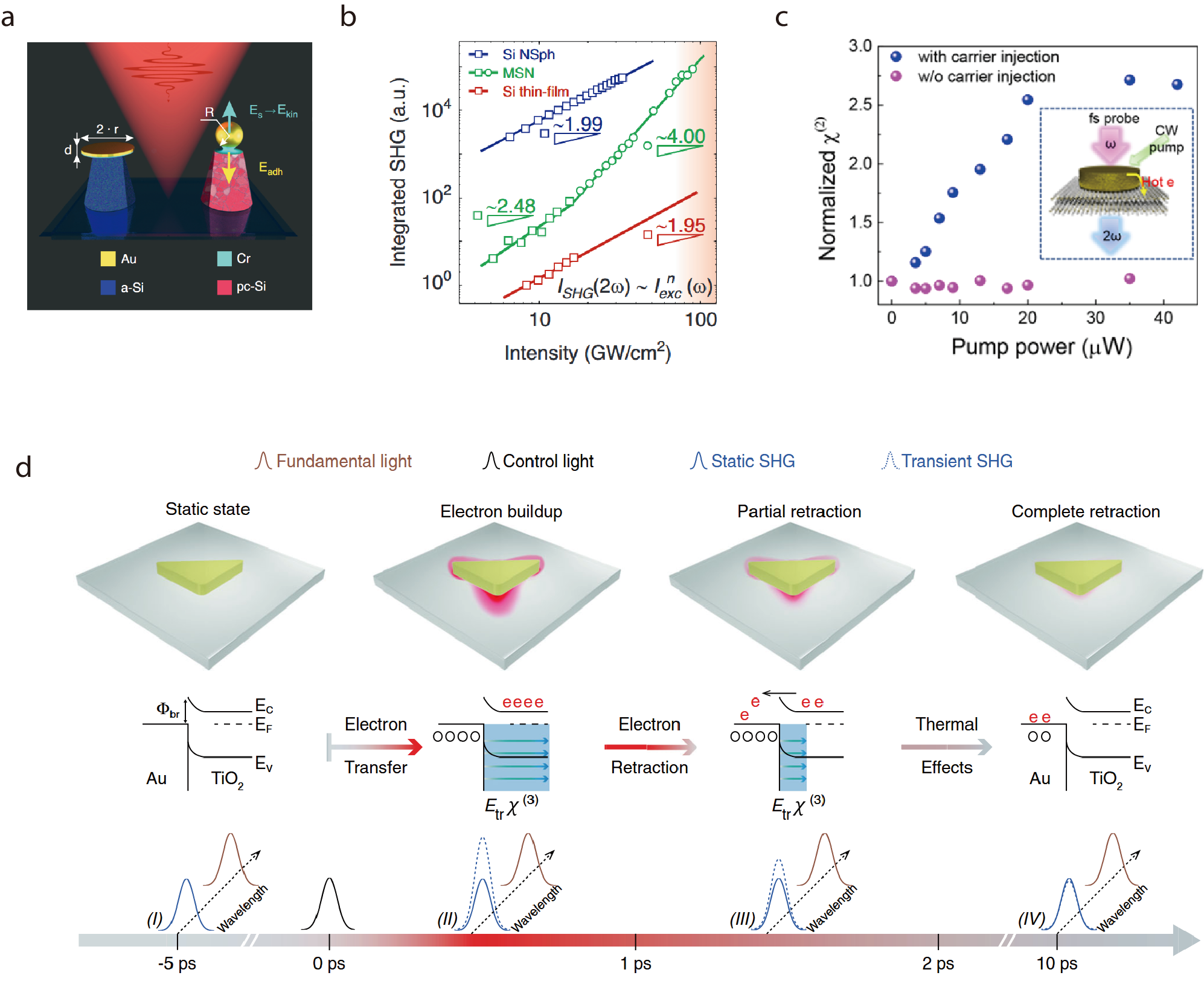}
  \caption{\textbf{Optical induced EFISH in heterogeneous structures.}
    a) Schematic of the hybrid MSNs.
    b) Evolution of the instantaneous slope $n$ of the SHG intensity versus excitation power in a log–log plot. Both the silicon nanostructured single particle and the thin film exhibit a quadratic dependence of SHG intensity on pump power, characteristic of conventional second-order nonlinear processes. In contrast, the optically induced EFISH process demonstrates a clear deviation from the quadratic behavior, indicating a non-quadratic power dependence. Reproduced with  permission.~\cite{2023LightSci.Appl.SunAllopticalGenerationStatic} Copyright 2023, Nature Publishing Group.
    c) Illustration of the plasmon induced hot electrons form Au nanobar transfer to bilayer $WSe_2$,which establish a nonstatic separation electric field at the interface perturbation the centrosymmetry and produce SHG. Reproduced with  permission.~\cite{2018NanoLett.WenPlasmonicHotCarriersControlled} Copyright 2018, American Chemical Society.
    d) Mechanism of inversion‑symmetry breaking by hot‑electron transfer: (top) plasmon excitation produces and injects hot electrons into the dielectric; (middle) charge separation across the Schottky junction creates a transient field $E_{\mathrm{tr}}$, converting the host’s $\chi^{(3)}$ response into an effective $\chi^{(2)}\propto E_{\mathrm{tr}}\chi^{(3)}$; (bottom) the combined effect of asymmetric electron injection and $E_{\mathrm{tr}}$ breaks the crystal symmetry, enabling ultrafast all‑optical control of second‑order nonlinear processes such as SHG. Reproduced with  permission.~\cite{2020Phys.Rev.Lett.TaghinejadTransientSecondOrderNonlinearb} Copyright 2020, American Physical Society. }
  \label{fig:boat5}
\end{figure}

\begin{figure}
  \includegraphics[width=\linewidth]{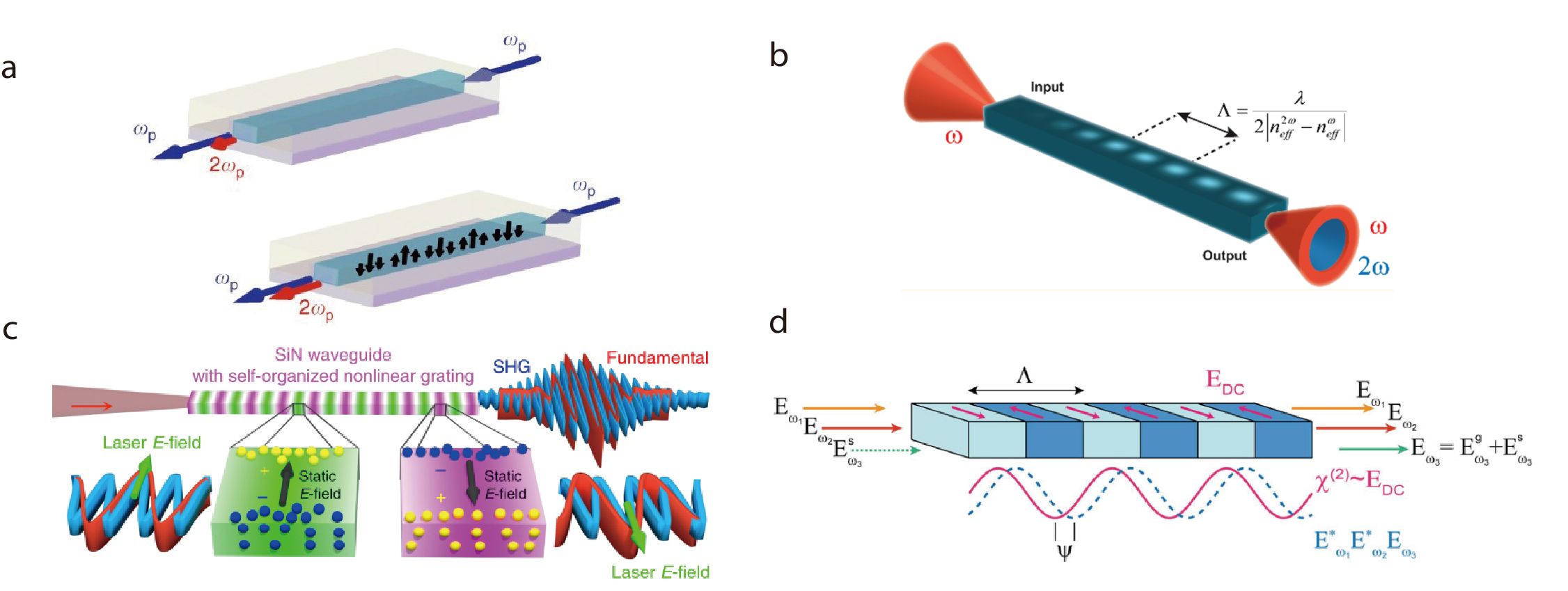}
  \caption{\textbf{Optically inscribed $\chi^{(2)}$ gratings in a SiN waveguide.}
    a) Illustration of $\chi^{(2)}$ grating inscription: upon irradiation by a pump at frequency $\omega_p$, a spatially periodic DC field develops with period which is the coherence length between the pump and second‑harmonic modes. Reproduced with  permission.~\cite{2017NatCommunBillatLargeSecondHarmonic} Copyright 2017, Nature Publishing Group.
    b) Charge‑separation mechanism: interference between the fundamental and second‑harmonic fields drives positive charges to one side of the waveguide and negative charges to the opposite side, establishing a static electric field that induces an effective $\chi^{(2)}$ grating. Reproduced with  permission.~\cite{2019Nat.PhotonicsHicksteinSelforganizedNonlinearGratings} Copyright 2019, Nature Publishing Group.
    c) Self‑grating formation rules: the inscribed grating follows the spatial modulation of the nonlinear polarization, producing a self‑organized quasi‑phase‑matched structure. Reproduced with permission.~\cite{2022ACSPhotonicsZabelichLinearElectroopticEffect} Copyright 2022, American Physical Society.
    d) QPM grating for SFG schematic: an optically inscribed grating with period $\Lambda$ can exhibit a relative phase shift $\phi$ between the $\chi^{(2)}$ modulation and the nonlinear polarization pattern of the interacting fields. Reproduced with  permission.~\cite{2022Laser&PhotonicsReviewsYakarGeneralizedCoherentPhotogalvanic} Copyright 2021, Wiley-VCH GmbH.}
  \label{fig:boat6}
\end{figure}

\begin{figure}
  \includegraphics[width=\linewidth]{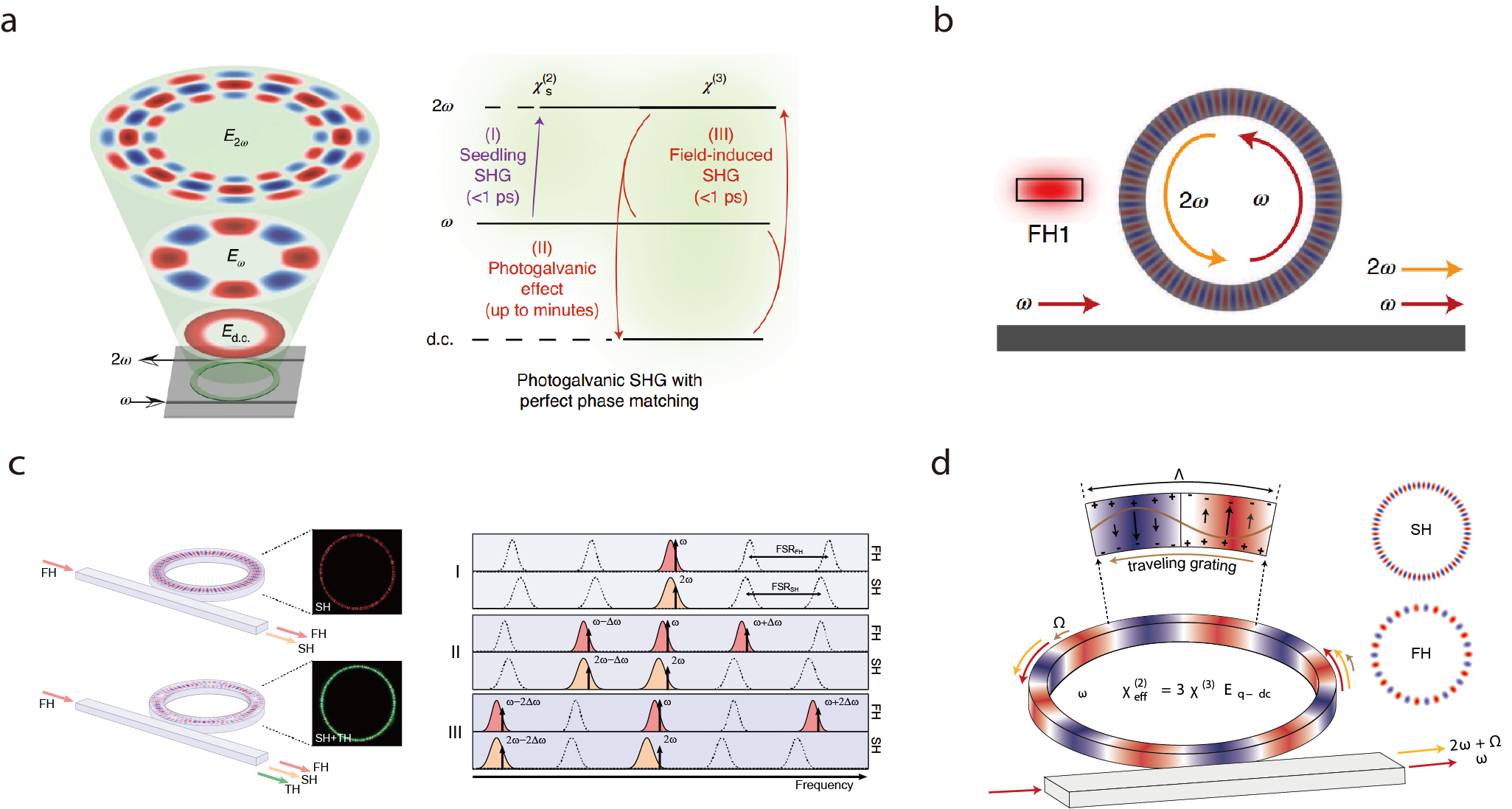}
  \caption{ \textbf{Optically inscribed $\chi^{(2)}$ gratings in a SiN microring}
    a) SHG via CPE with perfect phase matching. Illustration of the modes involved in perfect phase matching. Within the microring, the mode profiles of three interacting modes at $2\omega \text{, }\omega$ and d.c. frequencies are shown from top to bottom. Perfect phase matching induced uniform azimuthal distribution electrostatic field via CPE in a Si$_3$N$_4$ microring. Three‑step photogalvanic SHG build‑up: (I) a weak intrinsic $\chi^{(2)}_\text{s}$ seeds an initial SH field at $2\omega$; (II) interference of the $\omega$ and $2\omega$ fields generates a quasi‑static DC field via the coherent photogalvanic effect; (III) the DC field and pump produce additional SH via the d.c.\ Kerr effect, together amplifying the SHG signal. Reproduced with  permission.~\cite{2021Nat.PhotonicsLuEfficientPhotoinducedSecondharmonic} Copyright 2021, Nature Publishing Group.
    b) SHG with CPE ompensating momentum mismatching. Coupling of the fundamental TE mode (FH1, $\omega$) into a microresonator inscribes $\chi^{(2)}$ gratings defined by interference between FH1 and multiple SH TE modes (SH1–SH5). The azimutal distribution dc electrostatic field compensate the mismatch between FW and SH mode. Reproduced with  permission.~\cite{2022Nat.Photon.YuElectricallyTunableNonlinear} Copyright 2021, Nature Publishing Group.
    c) The SFG process via CPE compensating momentum mismatching in microring. Momentum conservation of SHG, SFG and cascade THG = SHG+SFG: quasi‑phase matching of SH, SF, and effective THG by the inscribed $\chi^{(2)}$ grating; SF‑coupled four‑wave mixing (FWM) produces sidebands $\omega\pm\Delta\omega$ (FH) and $2\omega-\Delta\omega$ (SH), tuning the comb free spectral range under pump detuning. Reproduced with  permission.~\cite{2022Sci.Adv.HuPhotoinducedCascadedHarmonic} Copyright 2022, American Association for the Advancement of Science.
    d) Spatiotemporal CEP: under doubly resonant conditions, the pump at $\omega$ inscribes a traveling $\chi^{(2)}$ grating oscillating at $\Omega$, generating SH at $2\omega+\Omega$; spatial and temporal modulation of $\chi^{(2)}$ governs momentum and energy conservation, illustrating the interplay between the coherent photogalvanic effect and EFISHG. Reproduced with  permission.~\cite{2025NatCommunZhouSelforganizedSpatiotemporalQuasiphasematching} Copyright 2025, Nature Publishing Group.}
  \label{fig:CPE_mirroring}
\end{figure}

\begin{figure}
  \includegraphics[width=\linewidth]{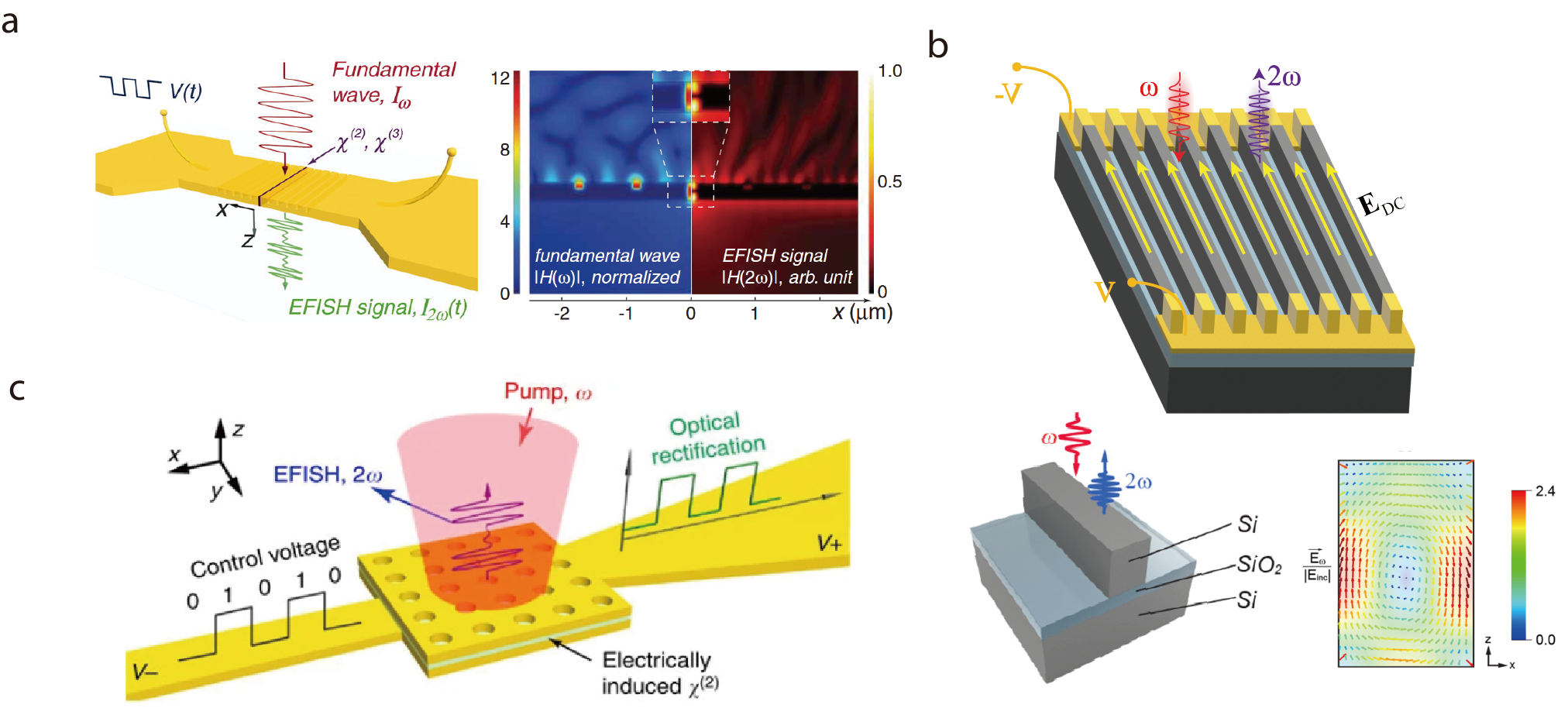}
  \caption{ \textbf{EFISH in metamaterials.}
    a) Schematic of a plasmonic EFISH device: a gold nanoslit resonator surrounded by a grating‑based optical nanoantenna, with both metallic parts wire‑bonded to external circuitry. Full‑wave simulations show the magnetic field distributions at the fundamental frequency $\omega$ and the EFISH signal at $2\omega$; the incident field is normalized to unity. Only half of the structure is displayed due to symmetry about $x=0$. Inset: detailed view of the plasmonic cavity. Reproduced with  permission.~\cite{2011ScienceCaiElectricallyControlledNonlinear} Copyright 2011, American Association for the Advancement of Science.
    b) Schematic of a Mie resonance EFISH device (upper), drawn based on Ref\cite{2019ACSPhotonicsKoshelevNonlinearMetasurfacesGoverned}. Unit‑cell schematic of a silicon metasurface with 2.4~fold field enhancement via MD resonance. Reproduced with  permission.~\cite{2019ACSPhotonicsKoshelevNonlinearMetasurfacesGoverned} Copyright 2019, American Chemical Society.
    c) Schematic of a metasurface EFISH device: a 50~nm perforated gold film separated from an unpatterned silver layer by a \SI{100}{\nano\meter} Al\textsubscript{2}O\textsubscript{3} spacer. The hole array has a diameter of \SI{186}{\nano\meter} and a square lattice period of \SI{370}{\nano\meter}. Both metallic films serve as electrodes when connected to external circuitry. Reproduced with  permission.~\cite{2014NatCommunKangElectrifyingPhotonicMetamaterial} Copyright 2014, Nature Publishing Group.}
  \label{fig:metaEFISH}  
\end{figure}

\begin{figure}
  \includegraphics[width=\linewidth]{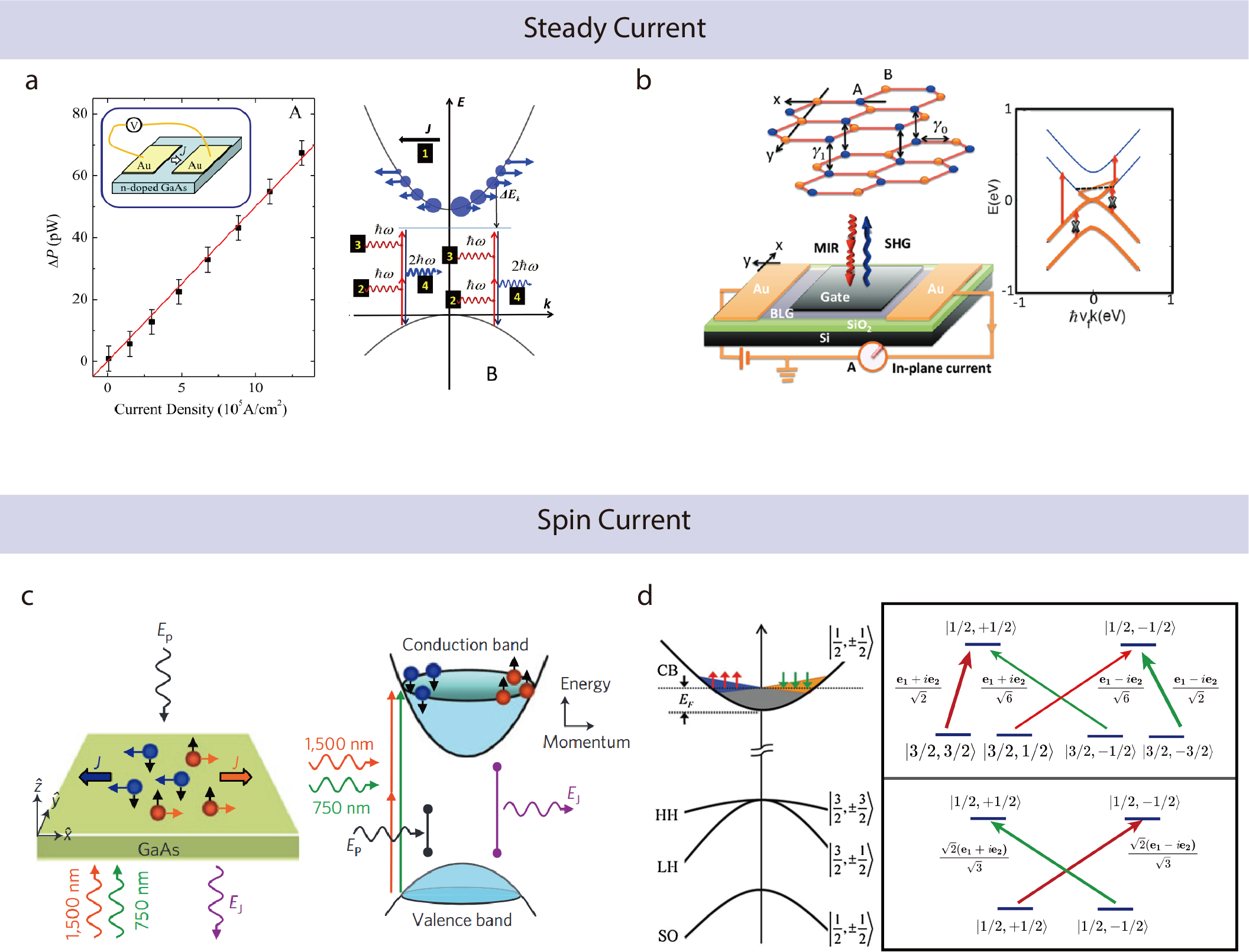}
  \caption{\textbf{Current induced SHG}
    a) Steady current induced SHG. Measured second‑harmonic signal as a function of charge current; inset: schematic of the device geometry. Microscopic model of the sequence of events in current-induced SHG. (1) DC current arises from an asymmetric carrier distribution in $k$-space; (2–3) virtual two‑photon transitions driven by fundamental photons, with transition matrix elements incorporating electron momentum and detuning $E_k$; (4) induced polarization at $2\omega$ emits the SH photon. Reproduced with  permission.~\cite{2012Phys.Rev.Lett.RuzickaSecondHarmonicGenerationInduced} Copyright 2012, American Physical Society.
    b) Dual‑gated bilayer graphene FET for SHG: (top) atomic structure of AB‑stacked BLG with interlayer ($\gamma_1$) and intralayer ($\gamma_0$) hopping parameters and coordinate axes; (middle) experimental schematic showing dual gates and source–drain bias, with in‑plane current along $-y$; (bottom) calculated BLG energy spectrum under bias, indicating Fermi level shifts, allowed (red arrows) and forbidden transitions, and occupied states (brown shading). Reproduced with  permission.~\cite{2012NanoLett.WuQuantumEnhancedTunableSecondOrder} Copyright 2012, American Chemical Society.
    c) Experimental setup for observing pure spin current (PSC)‑induced SHG: pump beams generate a PSC in the semiconductor sample, and the reflected second‑harmonic signal is detected. The GaAs sample is simultaneously illuminated by two laser pulses at 1550 and 750\,nm. The interference between the transition pathways driven by the two pulses (vertical red and green arrows in b) causes electrons with opposite spin orientations to be excited to energy states with opposite momenta (orange and blue spheres). As the two spin systems move along opposite directions, a pure spin current is formed. The nonlinear optical effect of the injected pure spin current is studied by detecting second-harmonic generation $E_\text{J}$ from a probe pulse $Ep$. Reproduced with  permission.~\cite{2010NaturePhysWerakeObservationSecondharmonicGeneration} Copyright 2010, Nature Publishing Group.
    d) Microscopic models for current induced sum‑frequency susceptibility: full eight‑band model showing the electron spin distribution with PSC and the relative dipole moments from the valence bands to the conduction band. Adapted  with  permission.~\cite{2010Phys.Rev.Lett.WangSecondOrderNonlinearOptical} Copyright 2010, American Physical Society.}
  \label{fig:current-induced}
\end{figure}

\begin{figure}
  \includegraphics[width=\linewidth]{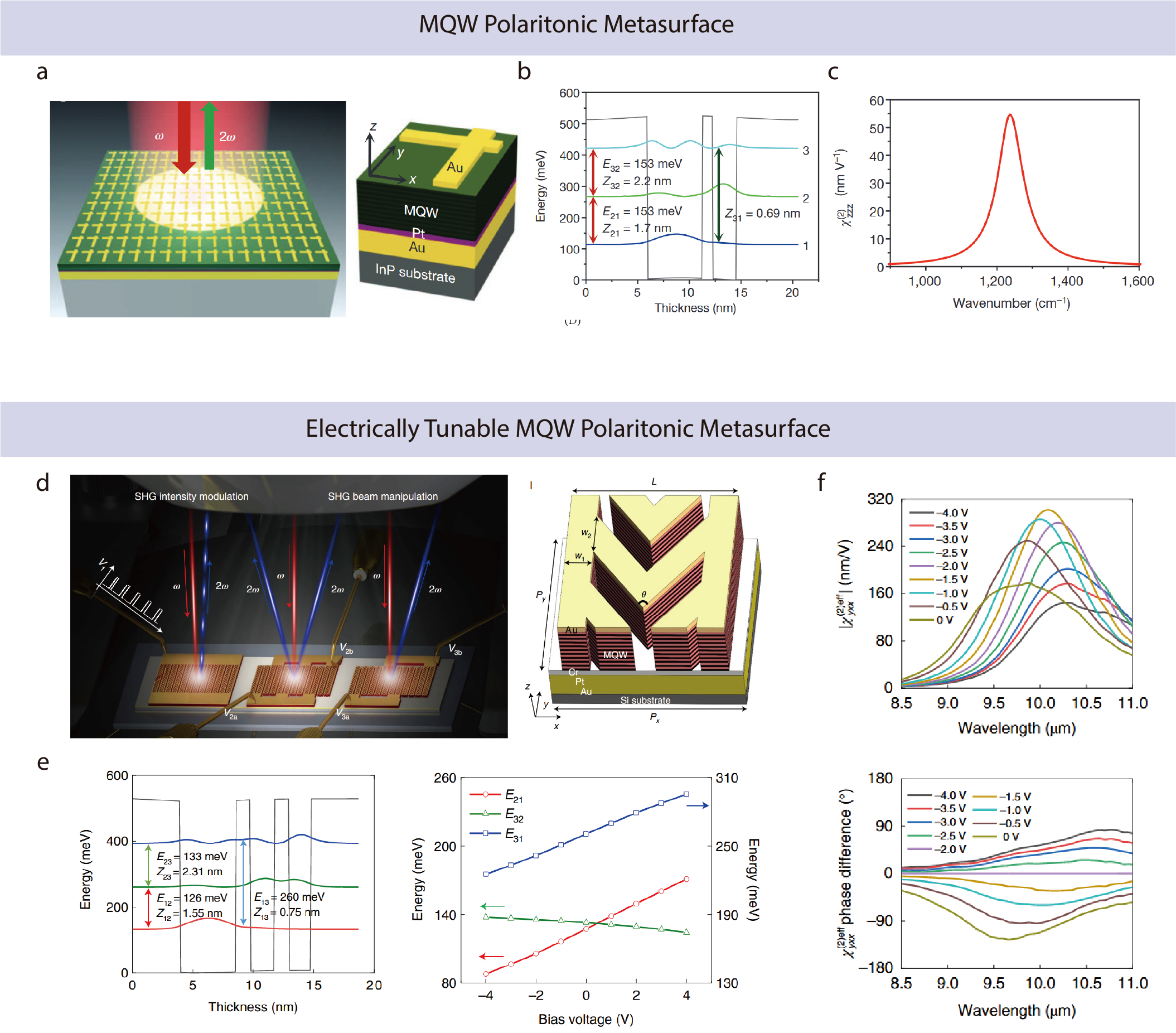}
  \caption{\textbf{Electrically tunable nonlinear polaritonic metasurface for SHG.}
    a) Schematic of the metasurface design.
    b) Conduction‑band diagram of one period of an In$_{0.53}$Ga$_{0.47}$As/Al$_{0.48}$In$_{0.52}$As coupled quantum well. Double‑headed red arrows indicate intersubband transitions, with energies $E_{21}$, $E_{32}$ and dipole moments $Z_{21}$, $Z_{32}$, $Z_{31}$.
    c) Intersubband nonlinear susceptibility of the structure in a) as a function of pump frequency for SHG. Reproduced with  permission.~\cite{2014NatureLeeGiantNonlinearResponse} Copyright 2014, Nature Publishing Group.
    d) Operation modes of the electrically tunable metasurface: SHG intensity modulation (left), beam diffraction (center), and beam steering (right). And schematic of the meta-atom unit structure.
    e) Conduction‑band diagram of the MQW unit under (left). The calculated IST energies under different bias  voltages (right).
    f) Calculated spectral dependence of the effective nonlinear susceptibility $\chi^{(2)}_{\text{eff},yxx}(V)$: magnitude (top) and phase (bottom) versus pump wavelength for various bias voltages. Reproduced with  permission.~\cite{2022Nat.Photon.YuElectricallyTunableNonlinear} Copyright 2022, Nature Publishing Group.}
  \label{fig:polaritonic-MS-for-SHG}
\end{figure}

\begin{figure}
  \includegraphics[width=\linewidth]{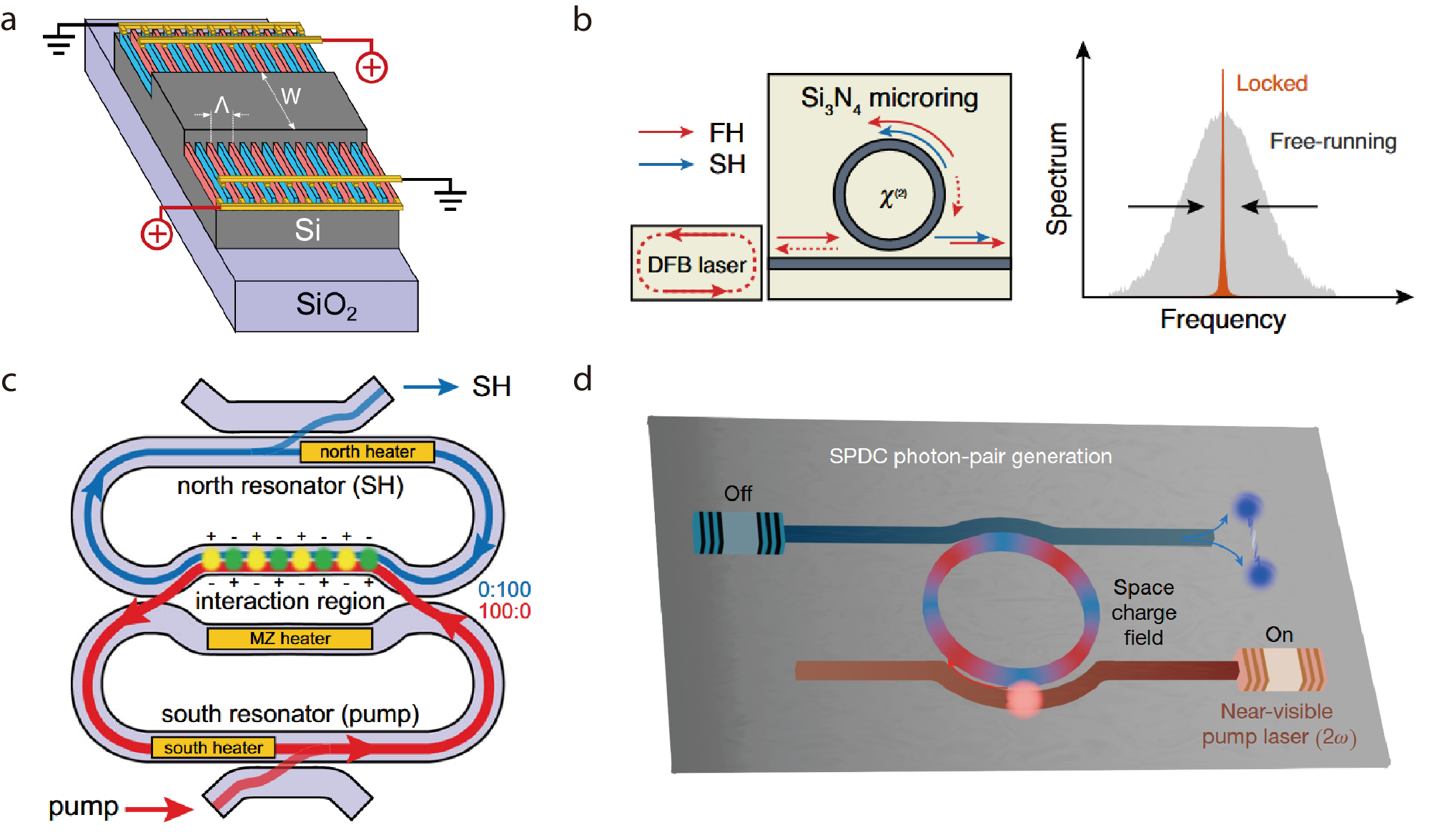}
  \caption{\textbf{Integrated nonlinear photonic devices.}
    a) Silicon rib waveguides with alternating p–i–n diode sections and electrical bias points for field‑induced nonlinear modulation. Reproduced with  permission.~\cite{2023OpticaHeydariDegenerateOpticalParametric} Copyright 2023, Optica Publishing Group.
    b) Self‑injection‑locking (SIL) scheme: a DFB laser injects light at the fundamental harmonic (FH) into a ring‑resonator bus waveguide (solid red arrow). Rayleigh backscattering (dashed arrows) feeds a fraction of the intracavity field back into the DFB cavity, dramatically narrowing the emission linewidth (lower panel). The resulting high‑coherence intracavity field drives the coherent photogalvanic effect to generate second‑harmonic light (blue arrow). Backscattered SH is omitted as it does not participate in SIL. Reproduced with  permission.~\cite{2023LightSciApplClementiChipscaleSecondharmonicSource} Copyright 2023, Nature Publishing Group.
    c) Ultrabroadband milliwatt-level resonant frequency doubling on a chip:Conceptual schematic of the device. The pump and SH circulate respectively in the south and north loops, sharing a fraction of the optical path where nonlinear interactions (CPE and SHG) occur. Reproduced with  permission.~\cite{2025NatCommunClementiUltrabroadbandMilliwattlevelResonant} Copyright 2025, Nature Publishing Group.
    d) Spontaneous parametric down‑conversion (SPDC) in a Si$_3$N$_4$ microresonator: after inscription of a space‑charge grating, a 780~nm pump is coupled into the resonator to produce near‑infrared entangled photon pairs at 1560~nm. Both pump and resonator are integrated on a semiconductor photonic chip. Reproduced with  permission.~\cite{2025NatureLiDownconvertedPhotonPairsa} Copyright 2025, Nature Publishing Group. }
  \label{fig:integrated-devices}
\end{figure}

\begin{figure}
  \includegraphics[width=\linewidth]{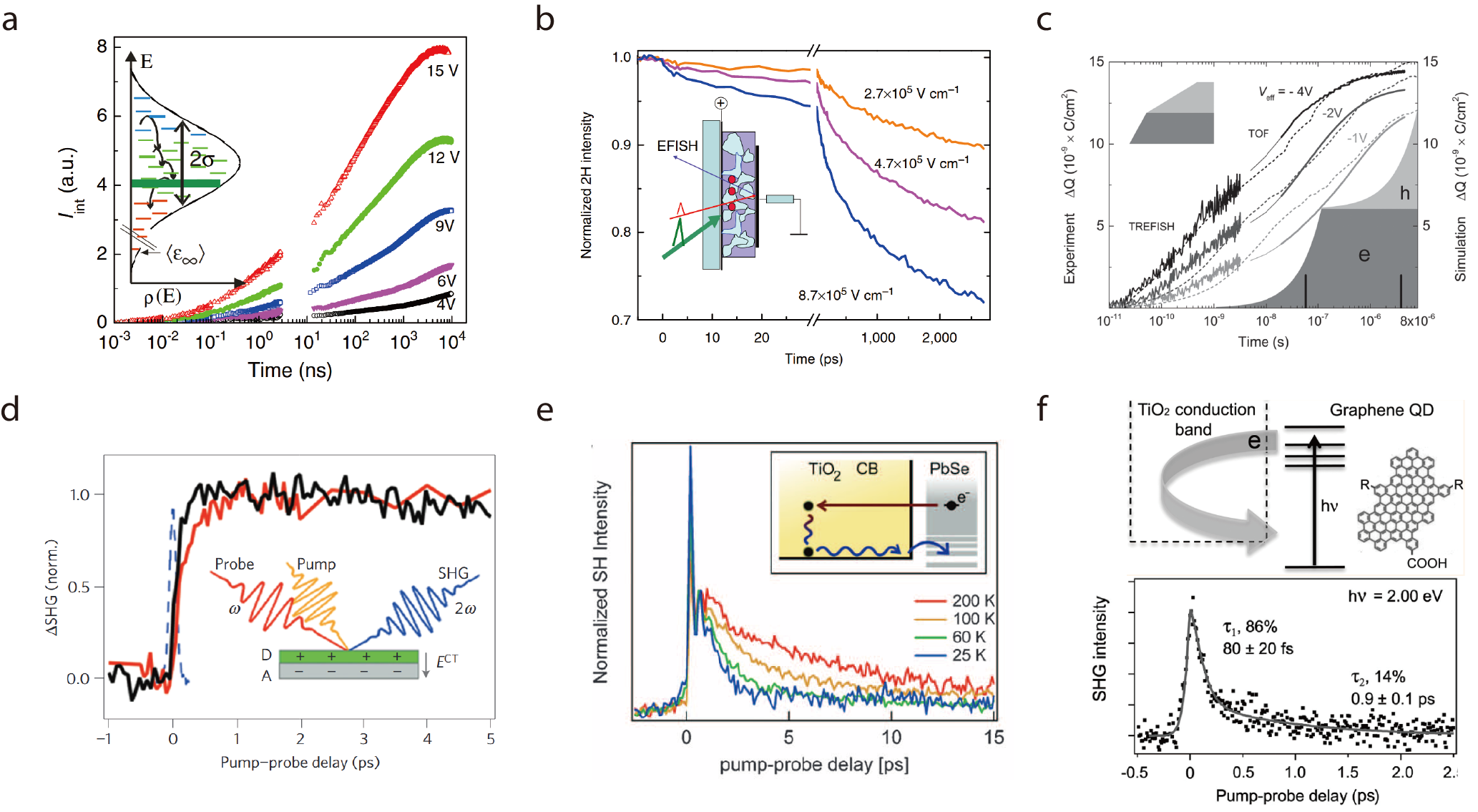}
  \caption{ \textbf{Time‑resolved EFISH application in Photovoltaic.}
    a) Integrated photocurrent from TREFISH and time‑of‑flight (TOF) measurements at various applied voltages after subtracting the exciton contribution $E_{\rm exc}(t)$. Inset: schematic density‑of‑states (DOS) in a disordered organic solid, showing the Gaussian DOS (width $2\sigma$), equilibrium carrier energy $E_{\rm eq}$, and transport energy (shaded bar). Reproduced with  permission.~\cite{2009Phys.Rev.Lett.DevizisUltrafastDynamicsCarrier} Copyright 2009, American Physical Society.
    b) Experimental kinetics of the second‑harmonic intensity under different electric field strengths. Reproduced with  permission.~\cite{2013NatCommunVithanageVisualizingChargeSeparation} Copyright 2013, Nature Publishing Group.
    c) Combined TREFISH/TOF transients (solid lines) and Monte Carlo simulation results (dashed lines) at various effective biases. The initial $\sim$20 ns (thinner lines) may be unreliable due to TOF limitations. Shaded regions indicate non‑dispersive extraction, with mean extraction times of $\sim$\SI{0.06}{\micro\second} (electrons) and $\sim$\SI{4}{\micro\second}~(holes) (vertical bars). Inset: same data on a linear time scale. Reproduced with permission.~\cite{2014AdvFunctMaterialsMelianasDispersionDominatedPhotocurrentPolymer} Copyright 2014, Wiley-VCH GmbH.
    d) TR‑SHG spectra of donor/acceptor bilayers: 3~nm CuPc on 20~nm C$_{70}$ (red, $h\nu=2.03$~eV) and 3~nm CuPc on 20~nm C$_{60}$ (black, $h\nu=2.02$~eV). Reproduced with  permission.~\cite{2013NatureMaterJailaubekovHotChargetransferExcitons} Copyright 2025, Nature Publishing Group.
    e) Temperature‑dependent decay of the pump‑induced SHG enhancement (normalized to the initial change) illustrating recovery rates. Inset: cartoon of ballistic electron injection (straight arrow), phonon scattering and polaron transport (wavy lines), and back‑transfer to the nanocrystal (curved arrow).Reproduced with  permission.~\cite{2010ScienceTisdaleHotElectronTransferSemiconductor} Copyright 2010, American Association for the Advancement of Science.
    f) Energy level diagram of the C132A graphene quantum dot (GQD) on TiO\textsubscript{2}(110): valence‑band maximum (VBM) and conduction‑band minimum (CBM). TR-SHG spectra (dots) for 0.5~ML C132A/TiO\textsubscript{2} at 300~K with pump photon energies of 2.00~eV (lower) and 2.10~eV (upper). Solid lines are biexponential fits; shaded regions indicate fit uncertainty. Probe photon energy: 1.53~eV; detected SHG at 3.06~eV.
    Reproduced with permission.~\cite{2013ACSNanoWilliamsHotElectronInjection} Copyright 2013, American Chemical Society. }
  \label{fig:solar-cells}
\end{figure}

\begin{figure}
  \includegraphics[width=\linewidth]{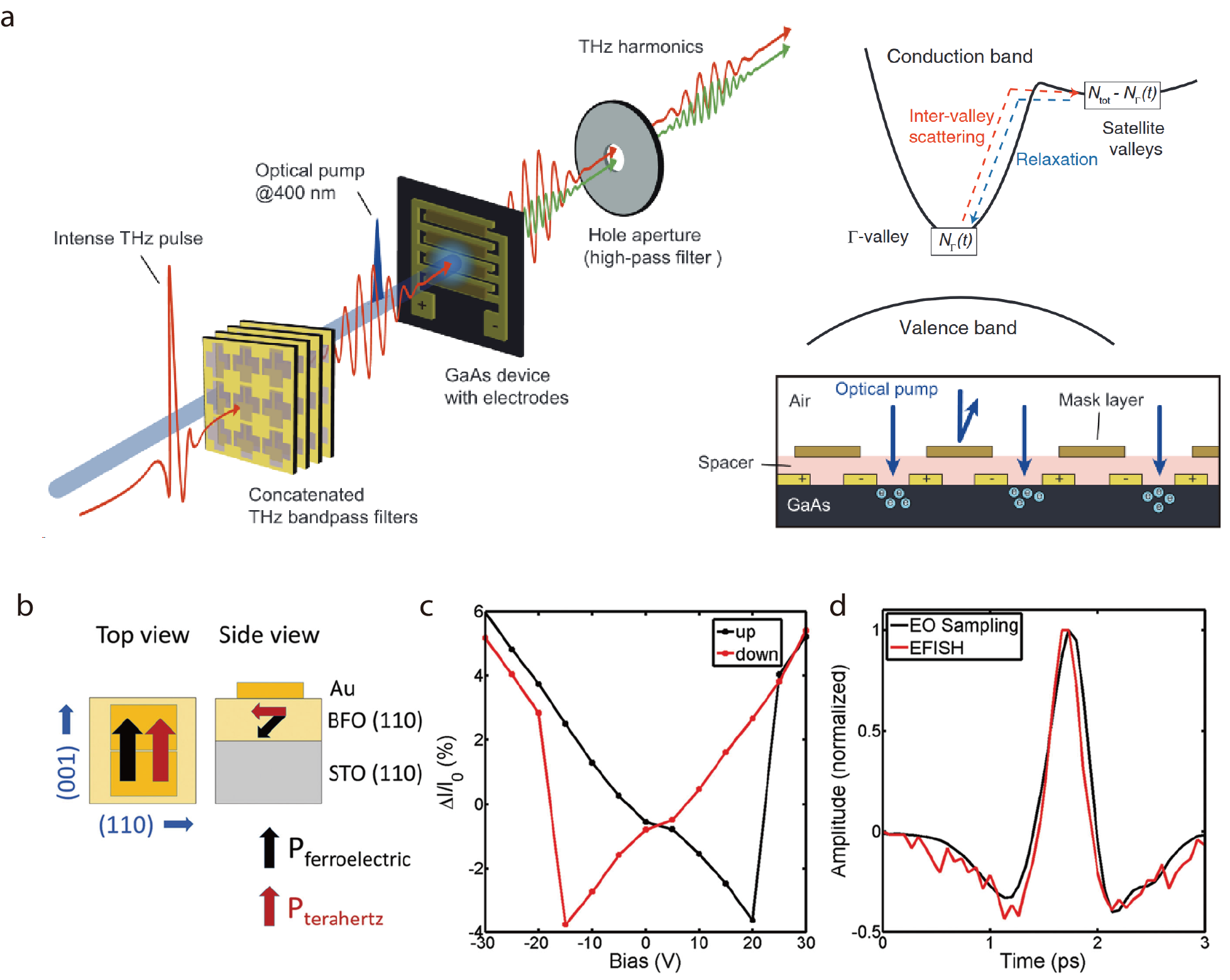}
  \caption{ \textbf{THz Application}
  a) Schematic of device and experimental setup. Schematic diagram of the band structure of GaAs and the intervalley scattering. In this diagram, $N\Gamma(t)$ and $N_{tot}$ denote the electron concentration in the $\Gamma$-valley and total electron concentration in the conduction band, respectively. Reproduced with permission.~\cite{2020AdvancedOpticalMaterialsLeeElectricallyControllableTerahertz} Copyright 2020, Wiley-VCH GmbH.
  b) Schematic showing sample electrode structure with incident THz pump pulse. 
  c) DC electrical biasing measurements with bias applied parallel to the ferroelectric polarization for a 2~$um$ gap on 60~$nm$ BFO film showing butterfly switching loop and switching fields on the order of $100\,kV/cm$. d) Low field THz EFISH measurement on the same BFO film shown in comparison to measured applied electric field profile measured by electro-optic sampling. Reproduced with permission.~\cite{2015AdvancedMaterialsChenUltrafastTerahertzGating} Copyright 2015, Wiley-VCH GmbH.}
  \label{fig:thz-application}
\end{figure}

\begin{landscape}
\begin{table}[htbp]
  \caption{Summary of Electrical Field Induced SHG System Performance Metrics}
  \centering\
  \begin{threeparttable}
  \begin{tabular}{@{}lllll@{}}
    \hline
    Material & Photonic Structure & Electrical Structure & Max SHG response~($\eta_{norm}$) & Modulation Depth \\
    \hline
    
    Lead titanate/Strontium titanate~\cite{2024NatCommunWangGiantElectricFieldinduceda}& Thin film & Metal electrode & $9.3\times10^{-11}\,W^{-1}$& $664\%\,V^{-1}$   \\
    
    Cadmium sulfide~\cite{2018NatCommunRenStrongModulationSecondharmonic}& Nanobar & Metal electrode & $4.4\times10^{-9}\,W^{-1}$& 200\%\,V$^{-1}$ \\

    Poly(9,9-di-n-dodecylfluorenyl-2,7-diyl)~\cite{2019LightSciApplChenGiganticElectricfieldinducedSecond}& Thin film & Metal electrode & $3.84\times10^{-13}\,W^{-1}$& 422\%\,V$^{-1}$ \\
    
    Poly(methyl methacrylate)~\cite{2011ScienceCaiElectricallyControlledNonlinear} & Metallic grating nanocavity & Metal electrode & $7.5\times10^{-15}\,W^{-1}$& 7\%\;V$^{-1}$ \\
   
    Aluminium oxide~\cite{2014NatureLeeGiantNonlinearResponse}& Metallic metasurface spacer &  Metal electrode & $2.67\times10^{-15}\,W^{-1}$& $9\%\;V^{-1}$ \\

    Silicon~\cite{2019ACSPhotonicsLeeElectricallyBiasedSilicon} & Metagrating & Metal electrode & -- & $1.2\,\text{V}^{-1}$\\
    Silicon~\cite{2017NaturePhotonTimurdoganElectricFieldinducedSecondorder}& Waveguide & p-i-n junction & $13\,\text{W}^{-1}$ & -- \\
    
    Silicon~\cite{2023OpticaHeydariDegenerateOpticalParametric} & Waveguide & p-i-n junction & $11.4\%\,W^{-1}$& -- \\
    
    Silicon nitride~\cite{2017NatCommunBillatLargeSecondHarmonic}& Waveguide & CPE  & $0.05\%\,W^{-1}$ &  \\
    
    Silicon nitride~\cite{2019Nat.PhotonicsHicksteinSelforganizedNonlinearGratings}& Waveguide & CPE  & $0.005\%\,W^{-1}$& -- \\
     
    Silicon nitride~\cite{2019Opt.Lett.GrassaniSecondThirdorderNonlinear}& Waveguide & CPE  & $0.008\%\,W^{-1}$& -- \\
    Silicon nitride~\cite{2020Photon.Res.NitissBroadbandQuasiphasematchingDispersionengineered}& Waveguide & CPE & $0.005\%\,W^{-1}$ & -- \\
    
    Silicon nitride~\cite{2020ACSPhotonicsNitissFormationRulesDynamics}& Waveguide & CPE & $0.31\%\,W^{-1}$ & -- \\
    
    Silicon nitride~\cite{2021Nat.PhotonicsLuEfficientPhotoinducedSecondharmonic}& Microring & CPE&$2500\%\,W^{-1}$\\
    
    Silicon nitride~\cite{2022Nat.Photon.NitissOpticallyReconfigurableQuasiphasematching}& Microring & CPE &$47.6\%\,W^{-1}$& -- \\
    
    Silicon nitride~\cite{2023LightSci.Appl.SunAllopticalGenerationStatic}& Microring & CPE  &$280\%\,W^{-1}$& -- \\
    
    Silicon nitride~\cite{2025NatCommunClementiUltrabroadbandMilliwattlevelResonant}& Microring & CPE  &$40\%\,W^{-1}$& -- \\
   
    Silicon nitride~\cite{2025NatureLiDownconvertedPhotonPairsa}& Microring & CPE  &$650\%\,W^{-1}$& -- \\

    Molybdenum ditelluride~\cite{2021NatElectronWangDirectElectricalModulation}& Monolayer & Metal electrode  & --&$3000\%\,V^{-1}$ \\

    Molybdenum disulfide~\cite{2017NanoLett.KleinElectricFieldSwitchableSecondHarmonic}& Bilayer  & Metal electrode  & --&$200\%\,V^{-1}$ \\

    Rhenium disulfide~\cite{2022ACSNanoWangElectricallyTunableSecond}& Trilayer  & Metal electrode  & --&$750 \%\,V^{-1}$ \\

    \hline
  \end{tabular}
  \begin{tablenotes}
    \item \textbf{Note:} The maximum SHG response is quantified by the normalized conversion efficiency $\eta_{\text{norm}}$, defined as $P_{2\omega}/P_{\omega}^2$ and $P_{2\omega}/(P_{\omega}P_{\omega,peak})$ under continuous‑wave and pulse laser excitation respectively, where $P_{\omega,\text{peak}}$ denotes the peak power of the incident pulse. The modulation depth is defined as ${\Delta I_{2\omega}(V_{\text{bias}})}/{V_{bias}I_{2\omega}(V_{\text{bias}}=0)}.$
\end{tablenotes}
\end{threeparttable}
\end{table}
\end{landscape}


\begin{figure}
  \includegraphics{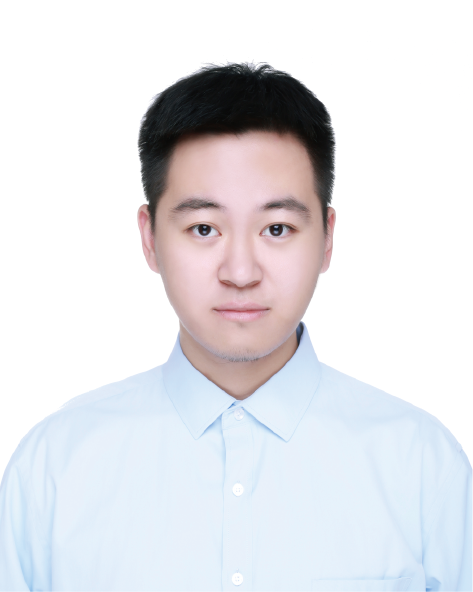}
  \caption*{\textbf{Hangkai Fan} is a Ph.D.\ student in Harbin Engineering University,China. Currently he is doing Ph.D. in Qingdao lnnovation and Development Base of Harbin Engineering University,China. His current research focuses on the nonlinear photonics in dielectric metasurface}
\end{figure}

\begin{figure}
  \includegraphics{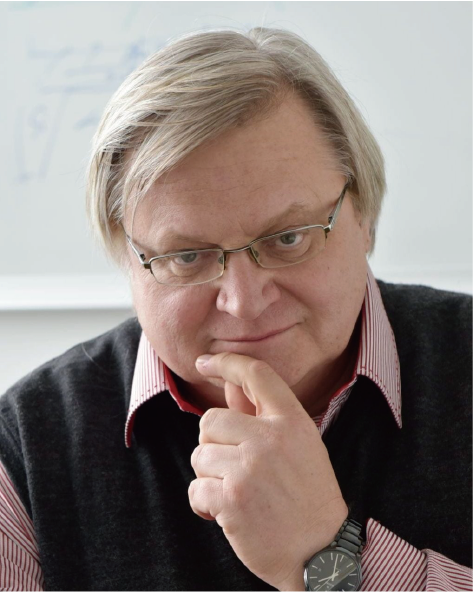}
  \caption*{\textbf{Yuri Kivshar} received a Ph.D.\ degree in theoretical physics in 1984 from the Institute for Low Temperature Physics and Engineering (Kharkov, Ukraine). From 1988 to 1993, he worked at various research centers in Europe and the USA. In 1993, he settled in Australia, where he established the Nonlinear Physics Center at the Australian National University; he is currently Head of the Center and Distinguished Professor. His research interests include nonlinear photonics, metamaterials, and nanophotonics. He has received numerous prestigious awards, including the Lyle Medal of the Australian Academy of Science, the State Prize in Science and Technology (Ukraine), the Lebedev Medal (Russia), the Harrie Massey Medal (UK), and the Humboldt Award (Germany).}
\end{figure}

\begin{figure}
  \includegraphics{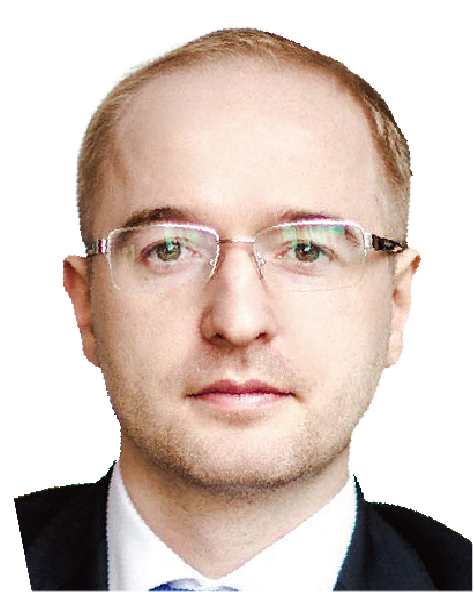}
  \caption*{\textbf{Andrey Bogdanov} received a Ph.D.\ degree in Semiconductor Physics from Ioffe Institute in 2012. He is currently working as a porfessor in Harbin Engineering University,China. He specialized in the study of resonant structures for the localization and control of light at the nanoscale.}
\end{figure}


\begin{figure}
\textbf{Table of Contents}\\
\medskip
  \includegraphics{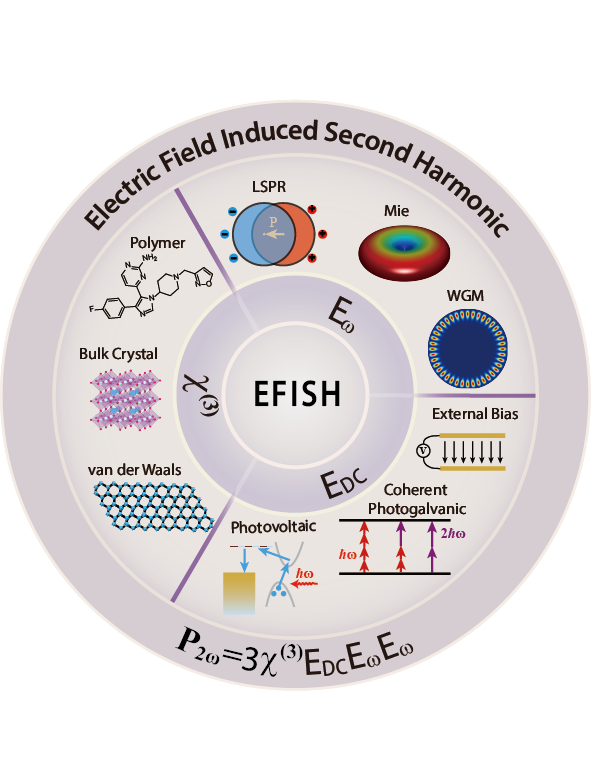}
  \medskip
  \caption*{Depending on the EFISH functional expression
\(\mathbf{P}_{2\omega}
=3\,\boldsymbol{\chi}^{(3)}:\mathbf{E}_{\mathrm{DC}}:(\mathbf{E}_{\omega}\otimes\mathbf{E}_{\omega}),\)
EFISH research can be classified into three main topics:(i) $\boldsymbol{\chi}^{(3)}$ engineering: selection of nonlinear materials, including donor–$\pi$–acceptor conjugated organic polymers, ferroelectric , and van der Waals material;(ii) $\mathbf{E}_{\mathrm{DC}}$ engineering: development of external‐electrode architectures and photoinduced asymmetric carrier distributions to tailor the static electric field profile;(iii)$\mathbf{E}_{\omega}$ engineering: design of high‐$Q$ optical resonators and metasurfaces to enhance the fundamental electric field distribution.}
\end{figure}

\end{document}